\documentclass[acmlarge]{acmart}
 \makeatletter
\newcommand{\myconfshort}{\acmConference@shortname}
\newcommand{\myconffull}{\acmConference@name}
\newcommand{\myconfdate}{\acmConference@date}
\newcommand{\myconfloc}{\acmConference@venue}
\AtBeginDocument{
  \fancypagestyle{firstpagestyle}{
    \fancyhead{}%
    \fancyfoot[C]{}%
  }
  \fancyhf{}
  \fancyhead[LO]{\@headfootfont\shorttitle}%
  \fancyhead[RE]{\@headfootfont\@shortauthors}%
  \fancyhead[LE]{\@headfootfont\footnotesize \myconfshort, \myconfdate, \myconfloc}%
  \fancyhead[RO]{\@headfootfont\footnotesize \myconfshort, \myconfdate, \myconfloc}%
  \fancyfoot[C]{}%
}
\makeatother
\acmBooktitle{\conffull\@ (\confshort), \confdate, \confloc}
\AtBeginDocument{%
  }
\copyrightyear{2026}
\acmYear{2026}
\setcopyright{cc}
\setcctype{by}
\acmConference[FAccT '26]{The 2026 ACM Conference on Fairness, Accountability, and Transparency}{June 25--28, 2026}{Montreal, QC, Canada}
\acmBooktitle{The 2026 ACM Conference on Fairness, Accountability, and Transparency (FAccT '26), June 25--28, 2026, Montreal, QC, Canada}
\acmDOI{10.1145/3805689.3812385}
\acmISBN{979-8-4007-2596-8/2026/06}
\usepackage{wrapfig}
\begin{document}

\title[Inside Baseball: ABS as an Object Lesson in Technological Rule Enforcement]{Inside Baseball: The Automated Ball-Strike System as an Object Lesson in Technological Rule Enforcement}

\author{Andrea Wen-Yi Wang}
\email{aww66@cornell.edu}
\affiliation{%
  \institution{Cornell University}
  \city{Ithaca}
  \state{NY}
  \country{USA}
}
\author{Waki Kamino}
\email{wk265@cornell.edu}
\affiliation{%
  \institution{Cornell University}
  \city{Ithaca}
  \state{NY}
  \country{USA}
}
\author{David Mimno}
\email{mimno@cornell.edu}
\affiliation{%
  \institution{Cornell University}
  \city{Ithaca}
  \state{NY}
  \country{USA}
}
\author{Karen Levy}
\email{karen.levy@cornell.edu}
\affiliation{%
  \institution{Cornell University}
  \city{Ithaca}
  \state{NY}
  \country{USA}
}
\author{Malte F. Jung}
\email{mfj28@cornell.edu}
\affiliation{%
  \institution{Cornell University}
  \city{Ithaca}
  \state{NY}
  \country{USA}
}
\begin{abstract}
Clearly-defined rules are often assumed to be straightforward to automate and evaluate. We challenge this assumption through an in-depth study of Major League Baseball’s (MLB) seven-year experimentation with the Automated Ball-Strike System (ABS). ABS is envisioned to call balls and strikes \textit{accurately}: a seemingly straightforward use of technology to objectively determine the distance between a pitch and the strike zone. Although the strike zone is an area clearly defined in the rulebook, it took MLB seven years to figure out how to automate calling balls and strikes with ABS, showing how even seemingly straightforward rules require a complex translation process to operationalize via technological systems. In this paper, we trace the design decisions that led to the current implementation of ABS.
Our case study reveals that ``distance'' exists even between a clear rule and its technological implementation. Using analytic frameworks from Science and Technology Studies (STS), we show that such distance exists because (1) historically, the ``ground truth'' of the strike zone is contested: the rule in practice has always reflected a hybrid between the rulebook definition and umpires' enforcement decisions;
and (2) the use of ABS is embedded in an existing eco-system, where the implementation of a technological enforcement system needs to balance multiple stakeholder values. This perspective challenges conventional evaluation paradigms that center on the distance between a formalized rule and its technological implementation, and instead calls for evaluating how such systems are \textit{experienced} in practice. Addressing this question requires in-depth social science approaches, contributing to ongoing conversations in FAccT about the implementation and evaluation of sociotechnical systems.
\end{abstract}

\maketitle
\section{Introduction}
\begin{quote}
    \textit{There is an old story about the three baseball umpires, discussing how they make calls. The first said, ``I call them as I see them.'' The second, ``I call them as they are.'' The third, ``They ain't nothin' until I call 'em.'' \footnote{This saying is often repeated as a joke; the third umpire's line is attributed to the legendary Hall of Fame umpire Bill Klem \cite{paumgarten2003-NoFlagPlay, archives1985-CloseCalls}.}
    }
\end{quote}

Technological systems are increasingly relied upon to make judgments and enforce rules. This reliance depends on a widely held belief that these systems offer a capacity for technical accuracy that can surpass that of people. To this end, there is an inherent appeal in relying upon technology for accuracy judgments, particularly when the rules to be enforced appear straightforward. In this paper, we examine an example of one such system in practice: Major League Baseball (MLB)'s Automated Ball-Strike System (ABS). Commonly referred to as the ``robot umpire'' by media \cite{stark2025-WhatWeLearned}, ABS is a system that uses ball-tracking technology to judge whether a pitch passes through the \textit{strike zone}. MLB began experimenting with ABS as a means of officiating the strike zone in 2019, and in 2026 will begin integrating the technology into every major league regular-season game. 

We find that baseball is an ideal case study to illustrate the complexity of seemingly straightforward judgments and their enforcement via AI-driven technology. 
Unlike fuzzy normative concepts such as ``fairness'' or ``risk'', the strike zone is clearly defined in the rulebook. ABS, therefore, is a technology that aims to automate a rule that doesn’t \textit{seem} to need a lot of interpretation or translation (is the ball in the strike zone or not?). However, it has taken MLB seven years to figure out how to integrate ABS into the game, showing that even these seemingly straightforward rules require complex translation when being automated, and that the translation itself is a messy, iterative and dynamic process involving negotiations and communications to align with stakeholder values.

Drawing on understanding from Science and Technology Studies (STS), we show that the needs for complex translation arise for two reasons. First, the ``ground truth'' for the strike zone is historically contested. The official rulebook clearly defines the strike zone, yet concurrently defers to umpires' judgments as the final arbiter. The strike zone rule, in practice, has always been a hybrid between the rulebook definition and umpire's enforcement decisions.
Second, enforcement technology doesn't exist in a vacuum, but rather is embedded in an existing eco-system of stakeholders and norms, all of which participate in shaping the design and use of the technology \cite{pinch1984-SocialConstructionFacts}. 
Value-sensitive design scholars have shown that technologies embody multiple, and often competing, values rather than purely technical objectives \cite{flanagan2008-EmbodyingValuesTechnology, flanagan2014-ValuesPlayDigital, friedman2019-ValueSensitiveDesign}. Building on this literature, we identify four values---technical feasibility, economic interest, integrity and stability, and gamesmanship---that collectively shape the ABS implementation of the strike zone rule. 

Our in-depth study of ABS contributes to a new understanding of the sources of ``distance'' between a seemingly clear rule and its technological implementation. We show that this distance arises not only from technical constraints, but also from the need to balance multiple stakeholder values and account for the broader social implications of the technology. Additionally, we argue that enforcement technologies are better understood \textit{as the rule} itself rather than as mere approximations of it. Treating technology as an approximation, centers evaluation on its deviation from the rule as ``ground truth,'' reflecting a typical engineering-oriented approach. Instead, we argue that it is also essential to evaluate how people would \textit{experience} the system as deployed, similar to how MLB evaluates different implementations across iterations of ABS. Such evaluation requires in-depth social science research and potentially new evaluation frameworks, contributing to ongoing conversations in FAccT that emphasize the importance of social science methods for studying and evaluating sociotechnical systems \cite{selbst2018-FairnessAbstractionSociotechnical}.

\section{Related Work}

FAccT and FAccT-adjacent scholarship have contributed to our understanding of rule/technology relationships in important ways. Much of this work focuses on how we interpret ``fuzzy'' rules or normative values---that is, rules that, on their face, give rise to multiple reasonable and potentially incompatible interpretations, like ``fairness,'' ``risk,'' or ``justice''---and operationalize those rules into mathematical or otherwise measurable definitions \cite{jacobs2021-MeasurementFairness, srinivasan2024-GeneralizedPeopleDiversity, poe2024-ConflictAlgorithmicFairness, balagopalan2022-RoadExplainabilityPaved}. For example, \citet{amoore2006-BiometricBordersGoverning} analyze border control technologies that classify ``risky'' bodies; \citet{koepke2018-DangerAheadRisk} study pretrial risk assessment algorithms designed to predict ``dangerousness,'' a concept that the authors argue is ill-defined in a legal sense; \citet{kleinberg2016inherent} examine how multiple ``fairness'' definitions can be incompatible in the criminal justice context; and so on. Scholars from STS, law, and the social sciences have examined the relationships between rules and technology in a variety of spaces, for example, in enforcing traffic laws \cite{rich2012-ShouldWeMake}, border control \cite{amoore2006-BiometricBordersGoverning, Josie2025-HowTechPowers}, criminal justice systems \cite{koepke2018-DangerAheadRisk}, policing \cite{egbert2021-CriminalFuturesPredictive}, and labor practices \cite{levy2022-DataDrivenTruckers}, among other areas.  

One common thread among much of this scholarship is its emphasis on the challenges of interpreting these ``fuzzy'' concepts in sociotechnical systems, therefore examining questions like: who performs the translation of normative concepts into technical operationalizations \cite{bovens2002-StreetLevelSystemLevelBureaucracies,koepke2018-DangerAheadRisk}? What is the translation process like and how does it represent (or fail to represent) particular interests \cite{egbert2021-CriminalFuturesPredictive,stevens2021-SeeingInfrastructureRace,kiviat2025-ExceptionsAlgorithmicAge}? How do we evaluate whether the translation is done accurately or not \cite{jacobs2021-MeasurementFairness}? A complementary line of research focuses on aspects related to the role of human judgment within these tech-mediated rule enforcement regimes, such as the degree of discretion exercised by street-level bureaucrats \cite{bovens2002-StreetLevelSystemLevelBureaucracies, alkhatib2019-StreetLevelAlgorithmsTheory}, and the amount of human oversight we desire over algorithmic systems \cite{green2022-FlawsPoliciesRequiring, wagner2019-LiableNotControl,jones2017-RightHumanLoop}. 

Relatively less emphasis in FAccT has been placed on questions that arise when we consider rules that don't seem to require as much interpretation or discretion, but \textit{appear} to be well-defined, straightforward, ``cut and dried'' specifications that merely need to be enforced to the letter. Here, we could consider paradigmatic examples of ``threshold-based'' rules, like the speed limit or constraints on driving with a particular blood alcohol level: the question of whether the rule was violated appears to be relatively straightforward and non-controversial (for instance, did the car exceed the speed limit or not?). The questions of how these seemingly straightforward rules get enforced by technology is less explored within the FAccT literature.

However, these kinds of rules \textit{were} taken up with interest by legal scholars in the 2000s, who questioned whether it would be desirable for technology to facilitate to-the-letter adherence to rules (i.e., ``perfect enforcement'') \cite{zittrain2008-FutureInternetAndHow, mulligan-PerfectEnforcementLaw, kerr2010-DigitalLocksAutomation}. For example, in his provocatively titled article ``Should We Make Crime Impossible?'' \citet{rich2012-ShouldWeMake} examined the normative questions surrounding an interlock device that aims to ``perfectly'' prevent drunk driving by stopping a car from starting when a certain amount of alcohol was detected on a person's breath. Much of this scholarship recognized that there might be good reason \textit{not} to favor perfect enforcement of law in all cases---including the possibility of productive disobedience, the need for ``human-in-the-loop'' judgment in exigent cases, and the dependence of social and economic life on some routine rule-breaking \cite{levy2022-DataDrivenTruckers, rich2012-ShouldWeMake}. However, the normative landscape that this scholarship typically explores assumes that technology would be capable of enforcing these rules in an accurate and precise way, leaving underexplored situations when technological systems reconfigure the straightforward rules themselves and create new, complicated enforcement realities. 

Our examination of ABS aims to bridge between FAccT's emphasis on translation, discretion, and oversight, on the one hand, and the legal literature's normative questions about the desirability of perfect enforcement, on the other. While all of these features (translation, perfect enforcement, discretion and oversight) are at play in the ABS case, none of them fully explain what we observe and document about the formulation and implementation of ABS. Unlike the translation literature, which focuses on rules that clearly require interpretation, we examine a seemingly clearly-defined rule. Unlike the ``perfect enforcement'' literature, which explores normative questions assuming technologies enforce rules to-the-letter, we identify a deliberate gap between a clear rule and its technological enforcement. And unlike the discretion and oversight literature, which typically centers on humans retaining final authority, we analyze a human-in-the-loop system in which the algorithm ultimately has the final say.
Therefore, our study on ABS pulls these several strands of scholarship on rules and technology together, and offers an object lesson for building insights about the relationship between rule and technology.

\subsection{Rules and Technology in Sports}
We chose sports as a strategic site through which to investigate technology and rule enforcement. In so doing, we build on a long line of work in both law and STS that takes sports seriously as a site of rule formation and technological enforcement. 
The metaphor of sports officiating as judgment is often used to minimize perceptions of a judge's role as a political or ideological actor.
Chief Justice John Roberts, in his confirmation hearing before the U.S. Senate's Judiciary Committee, compared a judge to a baseball umpire, ``calling balls and strikes'' on cases.  The metaphor suggests that the judge is merely a disinterested \textit{applier} of clear-cut and prespecified rules---nothing more \cite{fried2012-BallsStrikes, mckee2007-JudgesUmpires}.
Legal theorists also often use sports to illustrate theories about rules, enforcement, and discretion \cite{jones2018-SportingChancesRobot, lam2025-FewerRulesBetter, schauer1991-PlayingRulesPhilosophical}. Previous works in STS, human-computer interaction, and related fields have also identified the value of studying sports to study the interaction of technology, rules, and fairness \cite{kamino2025-MillionEyesRobot, shein2024-AIJudgingSports, collins2017-BadCallTechnologys}. Sports can function as a useful ``micro-legal'' environment \cite{kamino2025-MillionEyesRobot} in which we can develop accessible intuitions about the interactions of technologies and rules with normative values and social dynamics---acting as a prism to broader societal issues.

\section{Research Setting: Pitches, the Strike Zone, and the Automated Ball-Strike System in Baseball}
\begin{wrapfigure}[21]{r}{0.5\columnwidth}
  \centering
  \includegraphics[width=0.5\columnwidth]{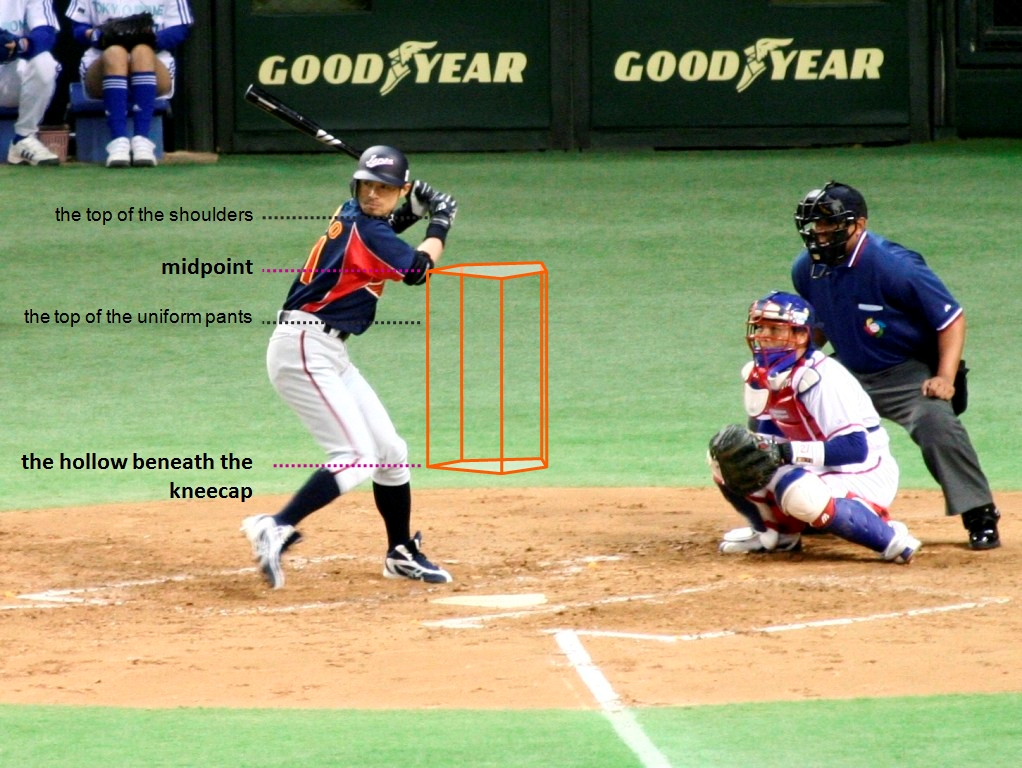}
  \caption{The official strike zone per the MLB rulebook, a three-dimensional prism illustrated with red lines. From left to right is the batter, catcher, and plate umpire. Photo by Mori Chan. Modified by the authors. Licensed under CC BY 2.0 \cite{wikimedia_morichan}.}
  \label{fig:strike-zone-def}
\end{wrapfigure}

A baseball pitch begins with one player (the pitcher) throwing a ball toward a teammate (the catcher). Between them stands an opposing player (the batter), who tries to hit the ball with a bat. As illustrated in Figure \ref{fig:strike-zone-def}, the pitcher's goal is to throw the ball so that it passes through a specific invisible box above home plate (the strike zone) while still being difficult for the batter to hit. On every pitch where the batter does not swing, the plate umpire must determine whether the pitch passes through the \textit{strike zone}: if it does, the umpires would call it a \textit{strike}, otherwise, a \textit{ball}.

Officially, per the MLB Rulebook, the strike zone is ``\textit{the area over home plate from the midpoint between a batter's shoulders and the top of the uniform pants -- when the batter is in his stance and prepared to swing at a pitched ball -- and a point just below the kneecap. In order to get a strike call, part of the ball must cross over part of home plate while in the aforementioned area}'' \cite{2025-OfficialBaseballRules}. Figure \ref{fig:strike-zone-def} provides a visual representation of the rulebook strike zone and its spatial relationship to the batter, catcher, and plate umpire. 

Starting in July 2019, MLB has experimented with the Automated Ball-Strike System (ABS), using pitch-tracking technology to automatically call balls and strikes in-game. They first experimented with the Trackman pitch-tracking system in their affiliated league, and later with the Hawk-Eye system \cite{Maaddi2019-RobotUmpiresDebut} in the minor leagues. The Hawk-Eye system tracks the movement of balls and players on the field with 12 high-speed cameras and computer vision algorithms \cite{jedlovec2020-IntroducingStatcast2020}. 

Implementing ABS is a complex and iterative process---MLB has spent seven years refining the system to settle on its implementation and facilitate buy-in from players, managers, umpires, and fans (see Table \ref{tab:abs-iterations}). One of the major hurdles has been to determine \textit{what} strike zone ABS should enforce \cite{AP2023-MLBTrackExpand, blum2021-AutomatedStrikeZone}, and how ABS should be used in a game. As we will see, the outcomes of these decisions were complex and far from inevitable, despite the seeming simplicity and straightforward nature of the strike zone rule.

In 2025, ABS was introduced in the major league for the first time in March during spring training, and again in July at the Major League All-Star Game. On both occasions, ABS was enforcing a strike zone that is \textit{not} the rulebook strike zone (see Table \ref{tab:abs-sz-vs-rule-book-sz}). Additionally, ABS was used \textit{selectively} as a ``challenge'' format, where human umpires are calling every pitch, but teams can appeal the umpires' calls using ABS. During these challenges, ABS will be used to arbitrate whether to overturn the human umpire's call. If ABS disagrees with the umpire, the umpire's call is overturned, and the challenge is \textit{successful}. Each team may challenge until it has two unsuccessful challenges in a game.
 MLB had also experimented with a \textit{full automation} format, in which ABS made every call and relayed the result directly to the plate umpire via an earpiece. Yet, a survey that MLB conducted in 2024 showed that players, managers and fans strongly prefer the \textit{challenge} system over \textit{full automation}. In fact, fans and players would rather have umpires making every call than have technology making every call. Among players and coaches, 54\% prefer challenges, 38\% favor human umpires, and only 8\% support full automation. Fans show a similar pattern: 47\% prefer challenges, 30\% favor human umpires, and 23\% support full automation.\footnote{Documents provided by MLB} In 2026, the ABS challenge system will officially be used in every major league regular-season game. 

\section{Research Methodology}
We take a multi-method ethnographic approach \cite{hammersley2019ethnography}, combining semi-structured interviews, field observation, and document analysis. The study ran between December 2024 and December 2025. Our study protocol was reviewed and approved by our institution's Institutional Review Board (IRB). 

\subsection{Data Collection}
\paragraph{Field study} To observe the technology in practice, we attended MLB Spring Training games as well as a minor league game where ABS was being tested and implemented. We also watched MLB's All-Star Game with ABS online. This fieldwork allowed us to witness firsthand how ABS Challenges function in live games and to observe interactions among players, umpires, ABS, and the fans' reactions.
In total, we visited four Spring Training games (including one without ABS for comparison), one Triple-A minor league game, and the All-Star Game (online). Moreover, during one of the Spring Training games, MLB allowed us to informally speak to ABS operators and an MLB official who showed us around the field and indicated, for example, the locations of the ABS cameras. 
Images and videos of ABS challenges, screenshots of surveys about ABS that fans receive at the end of the game were captured, and field notes included fans' reactions, and the flow of the ABS challenge practice in the stadium, mainly focusing on how ABS was experienced by fans present at the games.

In addition to baseball games, the second author attended and observed umpire training camps, which included shadowing a former MLB umpire coach and participating in both on-field and classroom lectures. This author attended two camps: the Southern Umpires Camp, a four-day program in Atlanta taught by former and current major and minor league umpires  \cite{atlanta}, and the official MLB Umpire Camp, a single-day camp run by MLB and held in Dallas \cite{dallas}. The first and second author also attended the 2025 Society of American Baseball Research Analytics Conference \cite{-SABRAnalyticsConference} to discuss and interact with analytical and technologically-oriented baseball fans. The first author visited the National Baseball Hall of Fame and Museum and the A. Bartlett Giamatti Research Center to explore historical documents and artifacts of baseball. 
\begin{table}[th]
    \centering
    \renewcommand{\arraystretch}{1.2}

    \begin{tabular}{@{}p{5mm}l p{5mm}l p{5mm}l@{}}
        \multicolumn{2}{@{}l}{\textbf{MLB}} &
        \multicolumn{4}{l@{}}{\textbf{Fans}}\\[2pt]

        & & \multicolumn{2}{l}{ } & \multicolumn{2}{l@{}}{ } \\[-6pt]

        E1 & Vice President, Umpire Operations &
        P1 & Male, Age 60s &
        P7 & Male, Age 30s\\

        E2 & Director, On-Field Strategy &
        P2 & Male, Age 20s &
        P8 & Male, Age 20s\\

        E3 & Senior Vice President, Engineering &
        P3 & Male, Age 20s &
        P9 & Male, Age 50s\\

        E4 & Senior Product Manager, Software Engineering &
        P4 & Male, Age 30s &
        P10 & Male, Age 70s\\

        E5 & Senior Manager, Data Operations &
        P5 & Female, Age 60s &
        P11 & Male, Age 60s\\

        E6 & Corporate Communications &
        P6 & Male, Age 30s &
        {} & {}\\

        E7 & Vice President, Head of Baseball Operations Counsel &
        {} & {} &
        {} & {}\\
    \end{tabular}
    \caption{Interview Participants}
    \label{tab:participant}
    \vspace{-10pt}
\end{table}

\paragraph{Interviews} 
We conducted 18 semi-structured interviews \cite{creswell2011-DesigningConductingMixed} between March 2025 and December 2025, including 7 with MLB executives involved in the design and implementation of ABS, and 11 with fans (see Table \ref{tab:participant}). 
The first and second author conducted these interviews via Zoom, phone calls, or face-to-face conversations. These interviews ranged from 45 to 60 minutes, with the exception of P10 (13 minutes) and E7 (120 minutes). In total, we have 840 minutes of interviews. All interviews were audio-recorded and transcribed for analysis. 

We contacted MLB to interview stakeholders involved in developing and implementing the ABS system. The On-Field Strategy team directed us to key personnel, and these interviews provided insights into their practices and perspectives. The MLB interviewees are identified with IDs E1--E7. We recruited eleven baseball fans through snowball sampling from our social network. Aged from their 20s to their 70s, these respondents started watching baseball regularly before their early teens. This resulted in different childhood exposures: older fans described attending ballparks, while younger fans described watching on TV. In addition to watching baseball, some fans engage with the sport in other ways, including playing baseball, organizing rotisserie (fantasy) leagues, or making a living by studying and writing about the game. Our interviews with fans started in late March 2025, after the ABS was first used at the major league level during Spring Training. All fans we interviewed had seen a game or video highlights involving the ABS prior to our interviews. Fan interviewees are identified with IDs P1--P11. 

\paragraph{Document review}
To historically contextualize the development of ABS, we conducted a supplementary archival review. Our sources include five books related to baseball rules recommended by an interviewee who is a baseball historian, around 40 media articles about MLB's experiments with ABS collected through web searches, and historical documents on the use of technology in baseball recommended by librarians in the Baseball Hall of Fame's Giamatti Research Center.

\subsection{Data Analysis Strategy}
We interpreted the data through thematic analysis \cite{braun2006-UsingThematicAnalysis}.
Themes were generated iteratively as we collected data. 
The first author created initial coding themes from the interview question list and by working through 2 MLB interviews and 2 fan interviews. Questions and codes were revised iteratively.
For example, we started the study with an exploratory question: ``What are people's perspectives about ABS and recent technological interventions in baseball?" After initial interviews with MLB stakeholders and fans conducted between March--May 2025, we noticed a key tension: the ABS system enforces a strike zone that differs from the official rulebook, yet fans praised ABS for making ``correct'' calls. The relationship between the rule and the enforcement technology became our analytical focus, and we began asking participants explicitly about their perspectives of this tension.

\section{The Contested Ground Truth of the Strike Zone}
Why is it hard to define the strike zone for ABS? By tracing the historical evolution of the strike zone rule, we observe that the ``ground truth'' of the strike zone has historically been contested, despite its seeming straightforwardness. 
Namely, despite the strike zone being clearly defined in the rulebook, the rule in practice has always been different---a hybrid between the rulebook definition and umpires' enforcement decisions. Further, recent technological developments have produced alternative strike zones \textit{visible} to spectators, which compete with umpires in defining what is seen as the rule's ground truth.

\subsection{Historical Roots of the Strike Zone}
When the game was first invented, baseball had no notion of balls, strikes, or a strike zone. The game's action centered on fielders and batters; pitchers' role was simply to deliver the ball for the batters to hit easily. But changes in strategy caused problems. In the era before electric lighting, games ended at predetermined times, either sundown or curfew. If a team was ahead, it could effectively suspend play to run out the clock by either having pitchers throw unhittable pitches or by having batters deliberately miss pitches.
In 1864, calling balls-and-strikes was developed as a mechanism to prevent players from exploiting this loophole, inducing pitchers to throw hittable pitches through ``balls'' and inducing batters to hit them with ``strikes'' \cite{hershberger2019-StrikeFourEvolution}.

Calling balls-and-strikes necessitated defining what is hittable and not, i.e. the \textit{strike zone}. Henry Chadwick, the ``Father of Baseball,'' who was most influential in shaping modern baseball rules, described the strike zone this way: 
\begin{quote}
    [The pitcher] must also deliver the ball ``for the striker''---that is, pitch him such balls as he is accustomed to strike at. Some batsmen liking low balls and some high. All, however, like the balls to come so that they can hit them with the bat within about six or eight inches of the end of it. And in order to get such balls they must stand so that the spot on the bat they want the ball to touch shall be over the home base. \cite{hershberger2019-StrikeFourEvolution}
\end{quote}
The definition of the strike zone changed several times over the next 150 years, for two main purposes: to make umpires' jobs easier, and to maintain the competitive balance between batters and pitchers. 
In the words of \citet{schauer1991-PlayingRulesPhilosophical}, the background justification of the strike zone rule is to define a location where a ball is \textit{hittable} by the batters. Initially, batters could call for a high or a low ball. The current version of the strike zone has been used since the 1996 season. 

\subsection{Three Umpires: Interpretations of the Strike Zone}
The gap between the formal, textual definition of the strike zone, and its background justification as a ``hittable'' area, can make it difficult to determine in practice what exactly is the strike zone (is it the area defined in the rulebook or is it the one that umpires call?). 
Unlike tennis court lines with visible white boundaries, a strike zone, while defined with reference to observable markers (belt, shoulders, plate), is not \textit{in itself} visible. On top of that, unlike tennis court lines that are unchanging, strike zones vary from batter to batter and even from pitch to pitch, depending on the batter's stance. 
Baseball rules designate the plate umpire's judgment as the final authority to determine whether a pitch is a ball or a strike. Balls-and-strikes calls by plate umpires cannot be argued or changed. In this sense, according to the rules of baseball, the plate umpire's decision constitutes the ground truth of the strike zone, as captured well in the joke about three baseball umpires at the beginning of the Introduction.

The joke is revealing in that it begins to clarify the tension behind the idea that the umpire's job is merely to enforce the rule of the strike zone as written---a tension which we will see emerge anew in the context of the ABS. The first umpire's view (``I call them as I see them'') emphasizes the potential imperfection of the (human) enforcement mechanism relative to the rule, suggesting that the human's perception drives the call of ball or strike, and leaving open, in theory, the possibility of distance between this perception and the rule itself (due to error, discretion, or for some other reason). The second umpire's phrasing (``I call them as they are'') collapses this gap between rule and enforcement, suggesting that the enforcement mechanism (the umpire himself) is a perfectly accurate instantiation of the rule. But the third umpire's phrasing goes further: rather than merely disagreeing over whether there is a distance between the rule and the enforcement mechanism, his phrasing (``they ain't nothin' until I call 'em'') suggests that the enforcement mechanism \textit{subsumes} the rule---there is no ``real'' strike zone, or ground truth, beyond what the umpire calls.

\subsection{The History of Making the Strike Zone Visible and Its Relationship to Umpires' Legitimacy}

\begin{wrapfigure}[19]{r}{0.65\columnwidth}
    \centering
    \includegraphics[width=\linewidth]{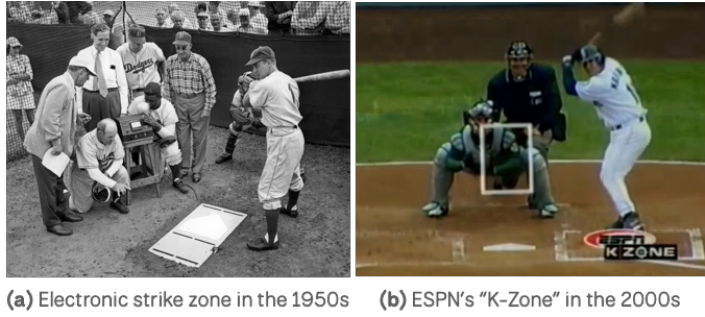}
    \caption{Making the strike zone \textit{visible}. (a) The Brooklyn Dodgers  experimented with electronic strike zones in the 1950s. Sports Studio Photos/Getty Images Sport via Getty Images \cite{2013-BrooklynDodgersTesting}. (b) ESPN began to overlay a computer graphic called the `K-Zone' since 2001 in their televised broadcast. YouTube video, posted by Detlef, 0:59 \cite{detlef2026-EvolutionESPNsKZone}.
    }
    \label{fig:strike-zone-device}
\end{wrapfigure}
Despite the rules designating umpires as the final arbiters of strike zones, there have been several previous attempts to visualize the strike zone with technology. In the 1950s, the Brooklyn Dodgers used ``mirrors, lenses and photoelectric cells'' to simulate a strike zone to train their players (Figure \ref{fig:strike-zone-device}) \cite{sandomir2016-ElectronicUmpireBaseball}.
The increased use of cameras in sports in the early 2000s marked a turning point for the strike zone and umpires. 
In 2001, strike zones started becoming widely \textit{visible} for TV audiences: ESPN began to overlay \textit{`K-Zone,'} a computer-generated graphic to visualize the strike zone, on their baseball broadcast to enhance the TV viewing experience (Figure \ref{fig:strike-zone-device}) \cite{sandomir2001-TVSPORTSYou, cafardo2011-Happy10thBirthday}. In the same year, MLB started using a first-generation camera-based pitch-tracking system, QuesTec, to train and evaluate umpires' balls-and-strikes calls \cite{dean-UmpiresTechYoure,Silver2003-ESPNcomMLBQuesTec}.
The year 2001, therefore, marked the start of an era in which there are \textit{other} strike zones competing with the one that umpires call.

The development of making the strike zone visible to spectators continued in later years. In 2006, ESPN started adding tracked pitch location on their computer-generated graphic of the strike zone \cite{cafardo2011-Happy10thBirthday}. In 2008, pitch-tracking systems were installed in all Major League ballparks. In 2015, MLB debuted its data platform \textit{Baseball Savant},\footnote{https://baseballsavant.mlb.com/} where it releases pitch-tracking data for every pitch in every game to the public. The data include pitch movement, \textit{and} the location of the strike zone for each pitch \cite{MLB-PitchtrackingEraGlossary}. Baseball enthusiasts and mathematicians developed metrics and algorithms to evaluate umpires' balls-and-strikes calls \cite{hunter2018-NewMetricsEvaluating}. One of the most popular websites, Umpire Scorecards,\footnote{https://umpscorecards.com/} generates reports of every Major League umpire's accuracy and consistency for every game.\footnote{Their report indicated that individual umpires accuracy ranges from 91\% to 96\% in 2025.} 
Neither the televised strike zone nor the strike zone location indicated by MLB data is the rulebook strike zone. Most notably, while the rulebook strike zone is a 3D volume, both strike zones illustrated on televised broadcast and reported by MLB data platform are a 2D plane. 
Therefore, it is best to understand these strike zones as \textit{other strike zones} rather than \textit{the official strike zone}.

While spectators have offered opinions about balls-and-strikes calls throughout the history of baseball, for the first time in baseball history, these \textit{other} strike zones allowed spectators to be confident about their own judgments about what a ball and a strike is: visually, through the depiction of the strike zone on televised baseball broadcasts, and statistically, through fine-grained evaluative data about human umpires' accuracy. These developments paved the way for automating balls-and-strike calls. 

\begin{table}
  \renewcommand{\arraystretch}{1.2}
  \begin{tabular}{cp{6cm}p{6cm}}
    \toprule
    Parameter & Rulebook Strike Zone & ABS Strike Zone\\
    \midrule
    Top  & The midpoint between a batter's
shoulders and the top of the uniform pants when the batter is in his stance and prepared to swing at a pitched ball & 53.5\% of the batter's standing height\\
Bottom & A point just below the kneecap when the batter is in his stance and prepared to swing at a pitched ball & 27\% of the batter's standing height\\
Shape & 3D box & 2D plane\\
    Placement & Over the home plate & Midpoint of the home plate \\
  \bottomrule
\end{tabular}
\caption{The comparisons between the official strike zone defined in the MLB rulebook and the strike zone enforced by ABS.}
\label{tab:abs-sz-vs-rule-book-sz}
\end{table}

\begin{table}
  \renewcommand{\arraystretch}{1.2}
  \begin{tabular}{p{5cm}p{9cm}}
    \toprule
    Parameter & Variations of ABS Implementation\\
    \midrule
    Width of ABS Strike Zone& 17\,in (2019) $\rightarrow$ 19\,in (2020) $\rightarrow$ 17\,in (2023)\\
    Top of ABS Strike Zone& Height-based${^*}$ (2019) $\rightarrow$ 56\% of height (2021) $\rightarrow$ 51\% of height (2022) $\rightarrow$ Stance-based (2023) $\rightarrow$ 53.5\% of height (2024)\\
    Bottom of ABS Strike Zone & Height-based${^*}$ (2019) $\rightarrow$ 28\% of height (2021) $\rightarrow$ 27\% of height (2022) $\rightarrow$ Stance-based (2023) $\rightarrow$ 27\% of height (2024)\\
    Shape \& Placement of ABS Strike Zone& 3D (2019) $\rightarrow$ 2D at front of the plate \& at 10 inches from the backtip of the plate (2021) $\rightarrow$ 2D at midpoint of the plate (2023)\\
    How much should ABS be used to enforce the strike zone? & Full automation (2019) $\rightarrow$ Challenge \& Full automation (2022) $\rightarrow$ Challenge (2024)\\
  \bottomrule
\end{tabular}
\caption{Multiple specifications of ABS strike zone that MLB tested throughout their seven-year experiments. Figuring out what strike zone should ABS enforce was one of the most challenging tasks in implementing ABS, as noted by several MLB officials over the years \cite{AP2023-MLBTrackExpand, glaser2021-DidPitchClocks, blum2021-AutomatedStrikeZone}. We compile the information from multiple media reports  \cite{cooper2025-DitchingMLBRule, Linbergh2021-MLBJustTried,stark2024-MLBJustTweaked,blum2021-AutomatedStrikeZone,blum2023-WhatStrikeBaseball,Blum2023-StrikeZoneTripleA,blum2025-RobotUmpiresAre}. Where reports disagreed on the timing of a change, we report the earliest year. $^*$We are unable to determine the exact setting for the 2019 height-based configuration for the top/bottom of ABS strike zone due to inconsistent reports.}
\label{tab:abs-iterations}
\vspace{-15pt}
\end{table}

\section{Embedded Automation: How Stakeholders and Norms Shape ABS Implementation and the Rule}
For ABS to be successful, it needs to be embedded into an eco-system of professional baseball with existing stakeholders and norms. Therefore, a mix of technical and social values collectively shape ABS implementation and the underlying rules. In this section, we articulate how several such values have shaped the implementation of ABS, including why and how it adopts a strike zone that differs from the one defined in the rulebook (Table \ref{tab:abs-sz-vs-rule-book-sz}), and why ABS is used selectively as a ``challenge'' format rather than on every call. 
In Section \ref{sec:dc-technical}--\ref{sec:dc-stability} we show how technical feasibility (\ref{sec:dc-technical}), economic considerations (\ref{sec:dc-economic}), and integrity and stability (\ref{sec:dc-stability}) collectively shape what strike zone ABS can automate. 
Then, in Section \ref{sec:dc-challenge}, we show how using ABS in a challenge format preserves gamesmanship, which is commonly regarded as the ``art'' of baseball. 

\subsection{From Stance-Based to Height-Based: Technical Constraints in Strike Zone Design} \label{sec:dc-technical}

Per the MLB rulebook, the vertical boundaries of the strike zone are determined by a batter's batting \textit{stance}. Yet, the ABS strike zone determines the vertical boundaries by a batter's \textit{height}, measured ``standing straight up without cleats'' \cite{Castrovince2025-ABSChallengeSystem}.
We identify the role of technical feasibility that forces the ABS strike zone to diverge from the rulebook strike zone. 

MLB had tested both stance-based and height-based approaches, oscillating between the two. In 2019, they started with a height-based approach with Trackman, the older tracking technology. After upgrading the tracking technology to the current Hawk-Eye system, MLB experimented with the stance-based strike zone in 2023, setting the bottom as ``the bend of the back knee'' and the top as the ``midpoint of hips + 8\,in adjustment (roughly 1 ball above the belt).'' Eventually, MLB settled on the height-based method due to concerns about ``data reliability and operational executions'' (E2). A document provided by MLB cited technical issues with the stance-based method including the lack of ``a measurement point between the hip joint and the shoulder.''

This back-and-forth demonstrates that, as a basic matter, \textit{technical feasibility} is one key constraint that can prevent enforcement technology from simply implementing the rules as written. Here, the ABS system could not consistently or predictably enforce the strike-zone rule with reference to the batter's stance, so the system instead defines the zone relative to the batter's height. 
Technical feasibility can be understood as a constraint on allowable error and unreliability insofar as the rule is instantiated in the enforcement technology. 

\subsection{Arriving at 53.5\% and 27\%: The Economics of Baseball in Strike Zone Design}\label{sec:dc-economic}

Once MLB decided on using a height-based method to set the vertical boundaries of the ABS strike zone, the next question is deciding on the numerical values in setting the top and bottom of the strike zone. The current ABS strike zone sets the top at 53.5\% of the batter's height, and the bottom at 27\% of the batter's height. 
MLB tested several values before arriving at these two exact numbers: the top ranging from 51\% to 56\%, and the bottom between 27\% and 28\% (see Table \ref{tab:abs-iterations}). 
Underlying these variations are economic considerations central to professional baseball: attracting and maintaining viewership. 

MLB described monitoring changes in game statistics, such as strike-out rate or walk rate, to inform how different numerical settings of vertical boundaries affect the competitive balance between offense and defense, especially considering that ``our fans want more action in the game'' (E2):

\begin{quote}
    One of our initial objectives in formulating the zone dimensions was to increase the percentage of balls put into play (i.e., reducing strike-outs and walks), aligning with fans' desires to have more action in the game. We tested multiple settings at the top and bottom as well as different center points, and found it difficult to achieve the goal of more balls put in play. As we shrunk the zone to reduce strikeouts, we saw an offsetting increase in walks.  The ultimate 53.5\%/27\% setting produced a slight decrease in the strikeout rate during testing as it yields a zone that is slightly smaller than the current called zone. (E2) 
\end{quote}

In this case, we see a different set of constraints at work. The league has an interest in maintaining a desirable amount of ``action'' (here, balls actually put into play). This has to do with increasing the number of moments of excitement in baseball, as well as considerations having to do with game flow (i.e., maintaining games of a reasonable length for purposes of increasing viewership, both in person and on television). The MLB has long experimented with reshaping the rules in the interest of maintaining games that are fun to watch and don't take too long---most recently, by introducing a pitch clock in 2023 to minimize the amount of downtime in the game. Here, the measurements of the top and bottom of the strike zone don't appear to arise from any kind of inherent ``first principles'' as to whether a ball pitched at 53\% versus 54\% of a batter's height should be considered a ball or a strike. Instead, the goal here is maintaining a desirable balance and degree of efficiency within the game; the rule is adjusted to encourage this.

\subsection{From 3D to 2D: Integrity and Stability in Strike Zone Design}\label{sec:dc-stability}
\begin{wrapfigure}[21]{r}{0.6\columnwidth}
    \centering
    \includegraphics[width=\linewidth]{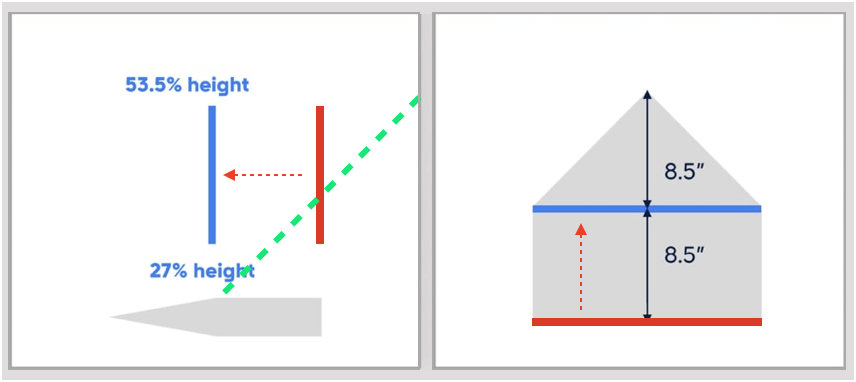}
    \caption{Moving the ABS strike zone from the front of home plate (red) to the midpoint of the plate (blue) helps maintain the integrity and stability of existing calling norms. The green dotted line illustrates the trajectory of a curveball, which would intersect the 2D strike zone at the front of the plate and therefore be called a strike under a front-plane configuration. However, umpires rarely call such curveballs strikes. Moving the 2D strike zone to the midpoint of the plate avoids calling these pitches strikes, thereby better aligning ABS outcomes with umpires’ established calling practices. Image adapted from \cite{Castrovince2025-MLBTestingAutomated}.}
    \label{fig:design-decision-abs-placement}
\end{wrapfigure}

Per the MLB rulebook, the strike zone is a 3D prism over home plate (Figure \ref{fig:strike-zone-def}). Yet, MLB reconfigures this definition for ABS through two related design choices: first, redefining the strike zone from a 3D prism to a 2D plane; and second, once a 2D plane is adopted, determining where that plane is positioned along the depth of home plate. The ABS strike zone was eventually defined as a 2D plane placed at the midpoint of home plate, illustrated as blue lines in Figure~\ref{fig:design-decision-abs-placement}. The differences in shape and placement allow MLB to maintain the \textit{integrity and stability} of the game.

When MLB first started testing ABS in 2019, they implemented a 3D strike zone following the rulebook. Yet players described the zone as ``unique'' and ``uncomfortable'' \cite{Mayo2019-MajorLeagueBaseball, Brandt2019-MLBsTopProspects, Dykstra2020-ToolshedFarmsPrep}.

Specifically, players frequently complained about how the zone judged curveballs \cite{Brandt2019-MLBsTopProspects}. Curveballs make drastic downward movement over the short distance of the depth of the home plate, illustrated as a green dotted line in Figure \ref{fig:design-decision-abs-placement}. Per the rulebook definition, a curveball that clips even a small portion of the 3D box over home plate officially constitutes a \textit{strike}. In practice, however, umpires typically call such pitches as \textit{balls}. 

MLB therefore adopted a 2D plane, effectively shrinking the size of the rulebook strike zone, in order to make it less likely for curveballs to be called ``strikes'' by ABS. The subsequent experiments on \textit{where} to place the 2D plane were also a calibration process in order to find a strike zone that ``feel[s] like the same zone hitters and pitchers have come to know in their time playing baseball'' \cite{stark2024-MLBJustTweaked}. 

E1 described how MLB arrived at the placement of the 2D plane through a process of aligning the system with umpires' calling norms:
\begin{quote}
Over the course of the experimentation, we are always asking umpires for feedback on, for example, is this a pitch that you would never call a strike? [And] feedback from umpires where they said, ``if I call that pitch (a strike), the dugouts are going to go crazy and \textit{nobody's going to believe it, and you're going to lose the integrity of the system}.'' That's how we locate the strike zone, moving it from when we originally started, at the very front of home plate towards the middle of the plate. (E1)
\end{quote}

In words of \citet{bowker2000-SortingThingsOut}, curveballs function as a ``borderland'' case: although formally classified as ``strikes'' under the rulebook, they are routinely treated as ``balls'' in practice, exposing the tension between the rulebook definitions and calling norms. MLB chooses for the ABS strike zone to align with existing calling norms for the integrity and stability of the game. Here, the technology surfaces the inherent gap between the ``law on the books'' and ``law in action'', a concept first described by legal scholar Ezra Pound in 1910 \cite{pound1910law}. Rules as written often don't align with rules as practiced. In opting to align ABS more closely with practice than with the written rule, MLB's design decision relied on social acceptability: the league needed the new technology not to be received as a disruption of existing norms in order for it to be accorded legitimacy by players, umpires, and fans.

\subsection{From ``Full Automation'' to ``Challenge'': Preserving the Art of Gamesmanship through Selective Enforcement} \label{sec:dc-challenge}
How much should rule enforcement rely on ABS versus human umpires? MLB has tested both full automation, in which ABS calls every pitch, and a challenge system, in which umpires call every pitch and players may appeal those calls to ABS a limited number of times. A MLB document reports that the ``challenge'' format is preferred among players, managers, and fans.\footnote{The aggregated attitude from umpires is unclear. Many MLB executives mentioned that some umpires would prefer full automation over the challenge format so they won't be scrutinized for every call. See Rob Manfred's interview \cite{schmidt2025-QARobManfred}.} In the 2026 major league regular season, each team will be allowed up to two unsuccessful challenges per game.\footnote{Teams would retain their challenge if ABS overturns the umpires' calls. Therefore, the total number of challenged calls could theoretically be more than four per game. A document provided by MLB shows that in 2024 Triple-A there were on average 3.9 challenges in a game, roughly 1\% of all pitches.} 
We identify that the selective enforcement of ABS preserves and contributes to the value of \textit{gamesmanship} in baseball.

The fallibility of human judgments has historically functioned as a strategic resource for players and managers. How to operate in the gray zone between cheating and fair play, to many baseball fans and players, \textit{is} the art of baseball \cite{zumsteg2007-CheatersGuideBaseball}. One strategy that frequently surfaces in the discussion of ABS is catcher's \textit{pitch framing}, where catchers use subtle body movements to bring borderline pitches back into the strike zone so umpires perceive them as \textit{strikes}. To outsiders, pitch framing may seem to be a form of cheating, yet to the baseball community it is valued as an ``art'' \cite{golen2023-RoboUmpsReach}.
A major concern with ``full automation'' is that framing will become a lost art. It will not only affect players' careers but also deprive baseball of a long-cherished aspect of the game \cite{booth2023-MLBCatchersWary, schmidt2025-QARobManfred}.
Yet with the challenge system, catchers' pitch framing continue to be essential. Some even argue that framing will be \textit{more important}, as good catchers try to not only influence the perception of umpires but also batters. 

The challenge format not only preserves pitch framing as a strategy, but also contributes to gamesmanship by making it part of the game to decide when to rely on automated enforcement (i.e., to invoke a challenge). Selective use of automated enforcement, therefore, preserves the unpredictability and hence the art of the game. In so doing, it also ensures that baseball retains its focus on human skill and strategy even in the face of technological enforcement.

\begin{figure}
    \centering
    \includegraphics[width=0.9\linewidth]{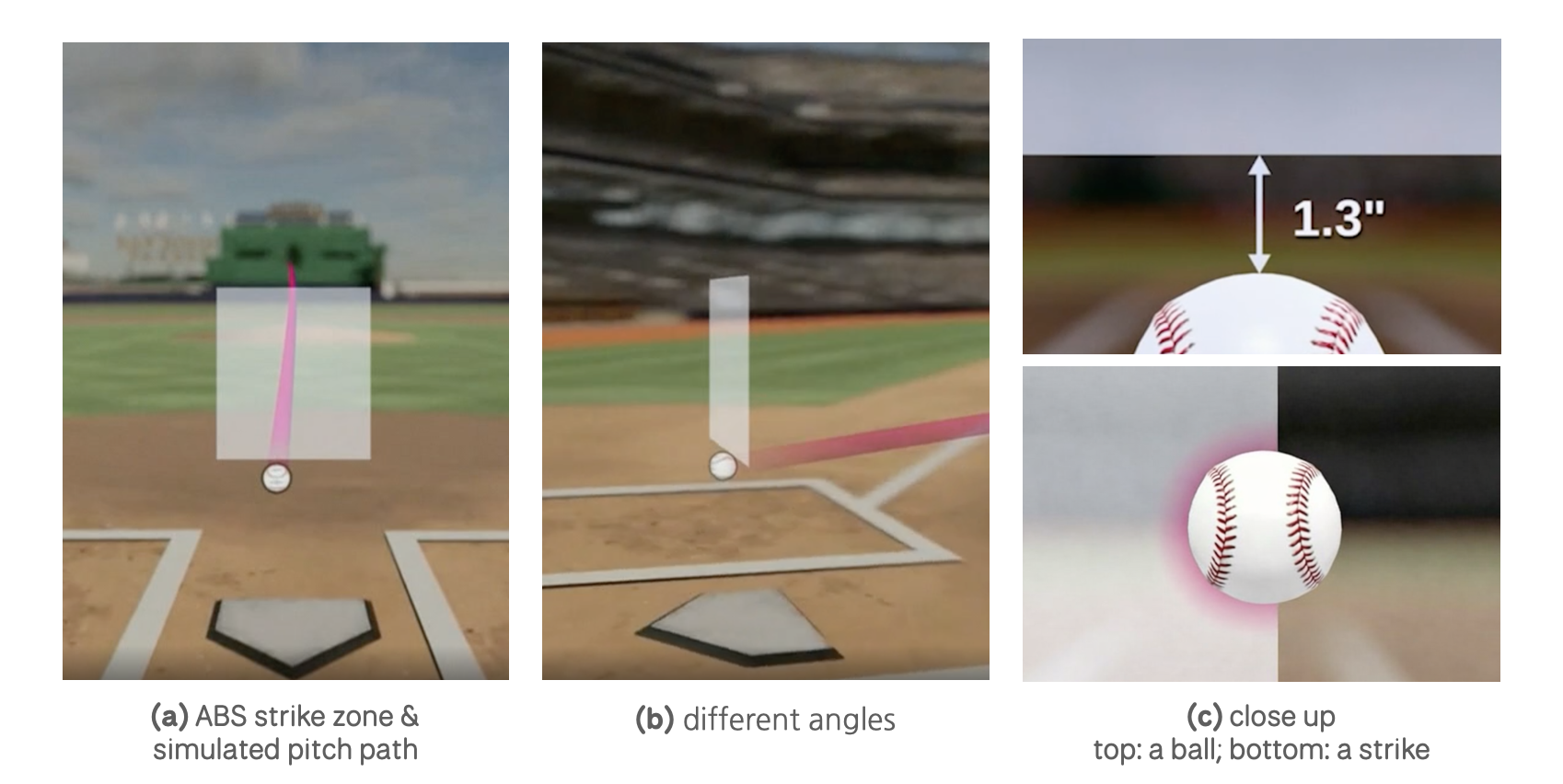}
    \caption{MLB designed a roughly 10-second video to play when players invoke ABS to challenge an umpire's call. This video started with (a) a simulated pitch trajectory and ABS strike zone from the viewpoint of an umpire, followed by (b) multiple angles to view the simulation, and ends with (c) a close-up of the ball and the boundary of ABS strike zone. For a `ball', the visualization additionally provides a numerical value showing how far outside the simulated baseball is from the ABS strike zone. Images adapted from \cite{Castrovince2025-ABSChallengeSystem}.}
    \label{fig:abs-viz}
\end{figure}

\section{Discussion}

\paragraph{On perceived accuracy}While ABS arose from the desire for calls to be made \textit{accurately}, what we have observed is that technical accuracy is not as important as the accuracy that is perceived and experienced by baseball stakeholders---and given the lack of a clear external ``ground truth,'' even describing ABS as ``accurate'' or ``inaccurate'' is a complicated proposition (accurate as compared to what?). However, the \textit{perception} of accuracy by stakeholders still matters.

MLB has designed a roughly 10-second video that displays the simulated pitch trajectory and the ABS strike zone, which is played when ABS is invoked to challenge an umpire's call. This visualization shows the simulated pitch trajectory and the ABS strike zone through multiple angles, before concluding with a close-up (Figure \ref{fig:abs-viz}). This visualization has been very successful in earning fans' trust in the system. Despite the lack of official performance metrics for ABS, fans perceive ABS as ``99\% accurate'' (P2), with a ``very slim'' margin of error (P3), citing the visualization as the source of confidence.\footnote{We were only able to identify two sources of official comments on the ABS system's margin-of-error. In 2021, Chris Marinak, MLB Chief Operations and Strategy Officer, stated that ABS accuracy is ``slightly plus or minus 0.1 inches'' \cite{blum2021-AutomatedStrikeZone}. A 2025 article in The New York Times, citing league officials, reports that ``the median error of the Hawk-Eye system is 0.17 inches'', and that MLB is ``99 percent confident it would be within 0.48 inches'' \cite{rosenthal2025-AutomatedBallStrikeSystem}.} For example, one fan described how ``the animated 3D graphic is able to see a single seam clip the zone'' (P2). In addition to using visualization to earn spectators' trust, MLB also found a strike zone that players, managers, and umpires could come to agreement upon through iteratively testing different versions of the ABS strike zone. In this way, MLB found a strike zone that could be implemented by ABS so stakeholders on the fields would \textit{experience} them as accurate and legitimate. 

The perceived accuracy of ABS, then, serves an important social function within the baseball eco-system: it serves as a purportedly objective ``outside observer'' to mediate disagreements. Historically, arguments involving balls-and-strikes calls could only result in players and managers being ejected from the field. With a strike zone that everyone can experience as accurate, ABS provides ``a sense of objectivity such that both teams and umpires could accept the use of ABS as the final arbiter when there is an argument'' (P4). One fan further described ABS as ``a robot god,'' that intervenes ``when two humans are arguing and shows what's actually right'' (P2).

These perceptions of accuracy are particularly interesting given the plethora of decisions that drive the design of the technology. As we have seen, those decisions have been driven by values beyond accuracy, including technical feasibility, economic interests, continuity with established enforcement norms, and the preservation of gamesmanship and the ``human element'' in the game. Therefore, we argue that the role of enforcement technologies should be understood beyond mere instantiations of a rule, and demonstrate that the values brought to bear within that technology's design exceed technical accuracy to include a number of social, economic, and cultural considerations. 

\paragraph{On the distance between a clear rule and its technological enforcement} As we have explored in this study, the translation from the strike zone, a rule clearly defined in the rulebook, to ABS, a technical implementation of that rule, is complicated despite the rule's seeming simplicity. The study leads to new understanding about the sources of ``distance'' between a seemingly clear rule and its technological implementation. First, the gap between ``rule on the book'' and ``rule in practice'' making it complicated to determine what the official strike zone actually \textit{is}. Technologists often conceive of rules on the books as deterministic and view technology as a means of overcoming human fallibility to instantiate the rulebook as written. However, the case for ABS shows that even a clearly defined rule resists such a simple characterization. Second, for ABS to be embedded smoothly into an eco-system with existing stakeholders and norms, a mix of technical and social values collectively shape the technological implementation of rules. Values such as technical feasibility, norms, and economic interests contribute to the ``distance'' between clearly-defined rules and their technological enforcement beyond the context of sports. Consider, for example, speed limit cameras, widely used in cities worldwide to automatically detect and penalize drivers who exceed the speed limit. While we might interpret their role as merely instantiating the pre-existing speed limit, many such systems actually only fine drivers for exceeding a speed at a threshold several miles \textit{above} the posted limit, in consideration of both technical feasibility concerns (having to do with calibration and accuracy thresholds) and norms (generally, law enforcement officers don't issue tickets for speeding one mile over the limit, and the social acceptability of speed cameras would plummet if these systems diverged from those enforcement practices). Economic interests also are primary considerations, as localities' interests in revenue generation have driven substantial debate about the ``correct'' number of speed violations, echoing our discussion about the top-and-bottom of the strike zone.

\paragraph{On evaluation} 
The typical frame for evaluating technology relies on a ground truth (the `rule'). Under this frame, the rule is pre-existing, and the goal of the technology is to instantiate that rule as closely and accurately as possible. 
However, we suggest that such an evaluation frame (i.e. trying to reduce the ``distance'' between rule and technology) obscures what is really experienced in the face of enforcement technology. 
In particular, we argue that \textbf{enforcement technology itself constitutes what the rule is}. 
Prior scholarship in the social sciences has made similar observations, documenting how technologies structure rule implementation and, in doing so, effectively become the rules themselves \cite{bovens2002-StreetLevelSystemLevelBureaucracies, koepke2018-DangerAheadRisk, egbert2021-CriminalFuturesPredictive, benjamin2019-RaceTechnologyAbolitionist, amoore2024-DeepBorder}. 
It is useful to go back to the three-umpire joke here. While popular narratives envision enforcement technology as the first umpire, faithfully enforcing a pre-existing rule (``I call them as they are''), in reality, it is like the third umpire: ``They ain't nothin' till I call 'em.'' Enforcement technology, like the third umpire, \textit{is} the rule. 
We find this reframing to be instructive for the FAccT community in raising the questions we ask about \textit{evaluation}, shifting from ``how faithfully does this implementation reflect the written rule?'' to ``how do people experience the enforcement technology? How have rules as experienced changed as they are implemented through technology?'' 

As an example from beyond the sports context, consider automated enforcement of social media content moderation policies. Algorithmic approaches to content moderation do more than simply translate written policies into code \cite{lessig2009code}. Rather, these technological implementations themselves \textit{are} the rules that users experience, leading savvy users to test the boundaries of what (say) a hate speech or toxicity detection algorithm will take down or devising ingenious ways to evade such detection. 

This perspective clarifies that when a machine learning or generative model is deployed as an enforcement mechanism, the de facto rule is the model itself. Regardless of how such systems are evaluated---through accuracy, precision, recall, or other metrics---their probabilistic nature means that enforcement is inherently variable. As a result, the rule, as experienced by users subject to these systems, is likewise probabilistic and unstable.

Shifting evaluation towards how people \textit{experience} technology has implications beyond enforcement systems. Consider, for example, educational technologies being deployed in K–12 classrooms, which are often rolled out at a much faster pace than that of ABS, leaving students feeling like guinea pigs \cite{maiberg2026-StudentsAreBeing}. While MLB spent seven years experimenting with its system and carefully assessing its potential effects, the same level of care has not been applied to many educational technologies. Standard evaluation approaches that measure how well a system produces a desired output can be conducted relatively quickly, whereas rigorous, in-depth studies of people’s experiences with these technologies require substantially more time and effort. Yet deploying technical systems in high-stakes contexts---arguably higher-stakes than baseball---demands at least the same, if not greater, level of evaluation, not only in technical terms but also in their social implications \cite{selbst2018-FairnessAbstractionSociotechnical}. 

\section{Limitations and Future Work}
We identify several limitations and future work directions. First, the gender distribution of fan interviewees in this study skewed male. While baseball viewership has historically been male-dominated, women's viewership has grown significantly in recent years \cite{kaplan2020-MLBTVRatings}. It would therefore be an interesting avenue for future research to investigate whether there are gendered differences in attitudes toward automation in rule enforcement in sports. Second, while this paper focuses primarily on the perspectives of MLB personnel and the history of the strike zone rule, future research could conduct in-depth studies with umpires and players to better understand how ABS influences their relationship with the game. Third, this research was conducted before ABS was formally implemented in MLB regular season games. Future work could examine how these dynamics evolve once ABS becomes an infrastructure of MLB games.
\section{Conclusion}
The Automated Ball-Strike System (ABS) is a technology developed to determine the location of a pitch relative to the strike zone. While the strike zone is an area clearly defined in the rulebook, it was far from easy to automate with technology. In fact, it took Major League Baseball (MLB) seven years of experimentation to decide what strike zone ABS should enforce.
The ABS case illustrates how enforcement technology surfaces the inherent gap between the ``law on the books'' and ``law in action.'' When MLB uses ABS to enforce the strike zone, it needs to consider the multiple stakeholder values that the strike zone implicates, beyond what's formalized on the rulebook. 
The need to balance multiple stakeholder values means that, for ABS to remain viable in baseball, it cannot simply conform to the formal rule.
ABS is valuable in showing how even seemingly very straightforward concepts (i.e., the strike zone) require translation when being automated, and that the translation itself is a messy process involving constant iterations, negotiations, and communications to align stakeholder values. Our study contributes to new understanding of the potential sources of ``distance'' between even a clear rule and its technological implementation.
Additionally, our study suggests that it is more instructive to treat enforcement technology \textit{as the rule} itself rather than an approximation of the rule. This shift in view is particularly relevant for FAccT when thinking about how we \textit{evaluate} enforcement technology. Rather than the typical evaluation setup where we treat the rule as ground truth and evaluate technology by comparing model output to that ground truth, it may be more meaningful to ask how users would experience this technology when deployed. Such questions require in-depth social science research and potentially new evaluation frameworks. 

\section{Generative AI Disclosure Statement}
All text in this paper was originally written by the authors. We used ChatGPT to lightly edit the writing, including to check grammar and to suggest ways to improve the fluency of the writing. We also searched for relevant papers to review using Ai2 Asta. 

\begin{acks}
We thank Dan Bateyko, Joshua Greenberg, Anurag Koyyada, and Travis Lloyd for the fun conversations we've had in the Rules and Technology seminar. We are grateful for the thoughtful inputs from John Conway, Lauren Klein, Daniel Mwesigwa, Ben Shestakovsky, Ranjit Singh, Malte Ziewitz, and the anonymous reviewers. We thank Major League Baseball and National Baseball Hall of Fame and Museum for the generosity of their time, materials, and expertise. We thank Eliza Bettinger, Iliana Burgos, Lencia McKee, Steven Streetman, and more from Cornell University Library for their generous support throughout the research and publishing process. 
This material is based upon work supported by the National Science Foundation under Grant No. IIS-2504533, and a Canadian Institute for Advanced Research Catalyst Grant. Any opinions, findings, conclusions, or recommendations in this material are those of the authors and do not reflect the views of the funding agencies.
\end{acks}

\bibliographystyle{ACM-Reference-Format}
\bibliography{1-zotero-references, 2-manual-references}

@book{2025-OfficialBaseballRules,
	address = {Chicago, IL},
	edition = {2025 edition},
	title = {Official baseball rules, 2025},
	isbn = {978-1-63727-803-1},
	language = {en},
	publisher = {Triumph Books},
	author = {Major League Baseball},
	year = {2025},
	note = {OCLC: 1520628966},
}

@book{friedman2019-ValueSensitiveDesign,
	title = {Value {Sensitive} {Design}: {Shaping} {Technology} with {Moral} {Imagination}},
	isbn = {978-0-262-35169-0},
	shorttitle = {Value {Sensitive} {Design}},
	doi = {10.7551/mitpress/7585.001.0001},
	language = {en},
	urldate = {2026-01-12},
	publisher = {The MIT Press},
	author = {Friedman, Batya and Hendry, David G.},
	month = may,
	year = {2019},
}

@article{archives1985-CloseCalls,
	title = {On {Close} {Calls}},
	url = {https://www.latimes.com/archives/la-xpm-1985-10-22-me-12377-story.html},
	language = {en-US},
	urldate = {2026-03-20},
	journal = {Los Angeles Times},
	author = {Archives, L. A. Times},
	month = oct,
	year = {1985},
	note = {Section: Sports},
}

@article{sandomir2016-ElectronicUmpireBaseball,
	title = {An {Electronic} {Umpire}? {Baseball} {Tried} {It} ({In} the 1950s!)},
	issn = {0362-4331},
	shorttitle = {An {Electronic} {Umpire}?},
	url = {https://www.nytimes.com/2016/05/01/sports/baseball/more-machine-than-man-an-electric-umpires-call-of-the-future.html},
	language = {en-US},
	urldate = {2025-12-12},
	journal = {The New York Times},
	author = {Sandomir, Richard},
	month = apr,
	year = {2016},
}

@article{booth2023-MLBCatchersWary,
	title = {{MLB} catchers wary of looming robo umps amid rules changes},
	url = {https://apnews.com/article/mlb-sports-minnesota-twins-rob-manfred-jose-trevino-063f72484e37a99e6ff428aaaf156d2d},
	language = {en},
	urldate = {2025-12-23},
	journal = {AP News},
	author = {Booth, Tim},
	month = feb,
	year = {2023},
}

@article{Brandt2019-MLBsTopProspects,
	title = {{MLB}'s top prospects deal with good, bad of 'robot' umpires},
	url = {https://apnews.com/mlbs-top-prospects-deal-with-good-bad-of-robot-umpires-2972799c8928483c92156c182909625b},
	language = {en},
	urldate = {2025-12-23},
	journal = {AP News},
	author = {Brandt, David},
	month = oct,
	year = {2019},
}

@article{AP2023-MLBTrackExpand,
	title = {{MLB} on track to expand robot umps to all {Triple}-{A} ballparks},
	url = {https://apnews.com/article/mlb-sports-rob-manfred-c4c8746bc3be2e932f109fc91cff6f81},
	language = {en},
	urldate = {2025-12-23},
	journal = {AP News},
	month = jan,
	year = {2023},
}

@article{blum2021-AutomatedStrikeZone,
	title = {Automated strike zone coming to minors but a while from {MLB}},
	url = {https://apnews.com/article/technology-mlb-baseball-rob-manfred-minor-league-baseball-fd64782047ad44f447e272375c712da5},
	language = {en},
	urldate = {2025-12-23},
	journal = {AP News},
	author = {Blum, Ronald},
	month = mar,
	year = {2021},
}

@article{blum2023-WhatStrikeBaseball,
	title = {What is a strike in baseball? {Robots}, rule book and umpires view it differently},
	shorttitle = {What is a strike in baseball?},
	url = {https://apnews.com/article/mlb-robot-umpires-strike-zone-40ec7285ae4d1ccaf2621adcb8d72b02},
	language = {en},
	urldate = {2025-12-26},
	journal = {AP News},
	author = {Blum, Ronald},
	month = jul,
	year = {2023},
	note = {Section: Sports},
}

@article{sandomir2001-TVSPORTSYou,
	title = {{TV} {SPORTS}; {You} {Have} {Entered} {The} {Strike} {Zone}},
	issn = {0362-4331},
	url = {https://www.nytimes.com/2001/07/19/sports/tv-sports-you-have-entered-the-strike-zone.html},
	language = {en-US},
	urldate = {2025-12-29},
	journal = {The New York Times},
	author = {Sandomir, Richard},
	month = jul,
	year = {2001},
}

@article{Josie2025-HowTechPowers,
	title = {How tech powers immigration enforcement},
	url = {https://www.brookings.edu/articles/how-tech-powers-immigration-enforcement/},
	language = {en-US},
	urldate = {2026-01-08},
	journal = {Brookings},
	author = {Stewar, Josie and Du, Michelle and Lee, Nicol Turner},
	month = oct,
	year = {2025},
}

@article{Mayo2019-MajorLeagueBaseball,
	title = {Major {League} {Baseball} tests robot umpires {Arizona} {Fall} {League}},
	url = {https://www.mlb.com/news/major-league-baseball-tests-robot-umpires-arizona-fall-league},
	urldate = {2026-01-12},
	author = {Mayo, Jonathan and Boor, William},
	month = sep,
	year = {2019},
}

@article{Castrovince2025-ABSChallengeSystem,
	title = {{ABS} {Challenge} {System} coming to {MLB} full time in '26},
	url = {https://www.mlb.com/news/abs-challenge-system-mlb-2026},
	urldate = {2026-01-08},
	journal = {MLB.com},
	author = {Castrovince, Anthony},
	month = sep,
	year = {2025},
}

@article{Castrovince2025-MLBTestingAutomated,
	title = {{MLB} testing automated ball-strike challenge system during spring games},
	url = {https://www.mlb.com/news/automated-ball-strike-calls-mlb-spring-games},
	language = {en},
	urldate = {2026-01-08},
	journal = {MLB.com},
	author = {Castrovince, Anthony},
	month = feb,
	year = {2025},
}

@article{Linbergh2021-MLBJustTried,
	title = {{MLB} {Just} {Tried} a {Bunch} of {Experimental} {Rules} in the {Minors}. {How} {Well} {Did} {They} {Work}?},
	url = {https://www.theringer.com/2021/10/21/mlb/experimental-rules-atlantic-league-robo-umps},
	language = {en},
	urldate = {2026-01-09},
	journal = {The Ringer},
	author = {Lindbergh, Ben and Arthur, Rob},
	month = oct,
	year = {2021},
}

@book{benjamin2019-RaceTechnologyAbolitionist,
	address = {Cambridge, England},
	title = {Race after technology: abolitionist tools for the new {Jim} code},
	isbn = {978-1-5095-2643-7},
	shorttitle = {Race after technology},
	language = {English},
	urldate = {2026-03-04},
	publisher = {Polity Press},
	author = {Benjamin, Ruha},
	year = {2019},
}

@article{rosenthal2025-AutomatedBallStrikeSystem,
	title = {Automated {Ball}-{Strike} {System} a major talking point ahead of possible 2026 {MLB} introduction},
	issn = {0362-4331},
	url = {https://www.nytimes.com/athletic/6366519/2025/05/19/automated-ball-strike-system-mlb/},
	language = {en-US},
	urldate = {2026-04-14},
	journal = {The New York Times},
	author = {Rosenthal, Ken and Stark, Jayson},
	month = may,
	year = {2025},
}

@article{blum2025-RobotUmpiresAre,
	title = {Robot umpires are getting their first {MLB} test during spring training},
	url = {https://apnews.com/article/mlb-robot-umpires-abs-9034454b5a795262bf97446ff38d0361},
	language = {en},
	urldate = {2026-04-22},
	journal = {AP News},
	author = {Blum, Ronald},
	month = feb,
	year = {2025},
}

@article{Blum2023-StrikeZoneTripleA,
	title = {Strike zone at {Triple}-{A} to get slightly bigger starting {Tuesday}},
	url = {https://apnews.com/article/robot-umpires-minor-league-strike-zone-2541180f8aaac1c79659a4e681e0a271},
	language = {en},
	urldate = {2026-04-22},
	journal = {AP News},
	author = {Blum, Ronald},
	month = sep,
	year = {2023},
}

@article{detlef2026-EvolutionESPNsKZone,
	title = {Evolution of {ESPN}'s {K}-{Zone} {Technology} [2001-2022]},
	url = {https://www.youtube.com/watch?v=AnHcyxU9PYA},
	urldate = {2026-05-10},
	journal = {YouTube},
	author = {{Detlef}},
	month = apr,
	year = {2026},
}

@article{cafardo2011-Happy10thBirthday,
	title = {Happy 10th birthday, {K} {Zone}},
	url = {https://www.espnfrontrow.com/2011/07/happy-10th-birthday-k-zone/},
	language = {en-US},
	urldate = {2025-12-29},
	journal = {ESPN Front Row},
	author = {Cafardo, Ben},
	month = jul,
	year = {2011},
}

@misc{2013-BrooklynDodgersTesting,
	type = {Getty {Images}},
	title = {Brooklyn {Dodgers}},
	url = {https://www.gettyimages.co.uk/detail/news-photo/brooklyn-dodgers-testing-new-automatic-umpire-at-vero-beach-news-photo/170185116},
	language = {en-gb},
	urldate = {2026-05-11},
	journal = {Getty Images},
	author = {Sports Studio Photos},
	month = jun,
	year = {2013},
}

@misc{glaser2021-DidPitchClocks,
	title = {Did {Pitch} {Clocks}, {Robo} {Umps} {And} {Other} 2021 {MiLB} {Rules} {Changes} {Work}?},
	url = {https://www.baseballamerica.com/stories/did-pitch-clocks-robo-umps-and-other-2021-minor-league-baseball-rules-changes-work/},
	language = {en-US},
	urldate = {2026-04-30},
	journal = {College Baseball, MLB Draft, Prospects - Baseball America},
	author = {Glaser, Kyle},
	month = sep,
	year = {2021},
}

@article{maiberg2026-StudentsAreBeing,
	title = {'{Students} {Are} {Being} {Treated} {Like} {Guinea} {Pigs}:' {Inside} an {AI}-{Powered} {Private} {School}},
	shorttitle = {Students {Are} {Being} {Treated} {Like} {Guinea} {Pigs}},
	url = {https://www.404media.co/students-are-being-treated-like-guinea-pigs-inside-an-ai-powered-private-school/},
	language = {en},
	urldate = {2026-03-23},
	journal = {404 Media},
	author = {Maiberg, Emanuel},
	month = feb,
	year = {2026},
}

@article{kaplan2020-MLBTVRatings,
	title = {{MLB} {TV} ratings increase over last season, led by women and youth},
	issn = {0362-4331},
	url = {https://www.nytimes.com/athletic/2012197/2020/08/21/mlb-tv-ratings-increase-led-by-women-and-youth/},
	language = {en-US},
	urldate = {2026-03-20},
	journal = {The New York Times},
	author = {Kaplan, Daniel},
	month = aug,
	year = {2020},
}

@article{paumgarten2003-NoFlagPlay,
	title = {No {Flag} on the {Play}},
	issn = {0028-792X},
	url = {https://www.newyorker.com/magazine/2003/01/20/no-flag-on-the-play},
	language = {en-US},
	urldate = {2026-03-20},
	journal = {The New Yorker},
	author = {Paumgarten, Nick},
	month = jan,
	year = {2003},
	note = {Section: dept. of super slo mo},
}

@article{jones2017-RightHumanLoop,
	title = {The right to a human in the loop: {Political} constructions of computer automation and personhood},
	volume = {47},
	issn = {0306-3127},
	shorttitle = {The right to a human in the loop},
	url = {https://doi.org/10.1177/0306312717699716},
	doi = {10.1177/0306312717699716},
	language = {EN},
	number = {2},
	urldate = {2026-03-19},
	journal = {Social Studies of Science},
	publisher = {SAGE Publications Ltd},
	author = {Jones, Meg Leta},
	month = apr,
	year = {2017},
	pages = {216--239},
}

@article{wagner2019-LiableNotControl,
	title = {Liable, but {Not} in {Control}? {Ensuring} {Meaningful} {Human} {Agency} in {Automated} {Decision}-{Making} {Systems}},
	volume = {11},
	copyright = {© 2019 The Authors. Policy \& Internet Published by Wiley Periodicals, Inc. on behalf of Policy Studies Organization.},
	issn = {1944-2866},
	shorttitle = {Liable, but {Not} in {Control}?},
	url = {https://onlinelibrary.wiley.com/doi/abs/10.1002/poi3.198},
	doi = {10.1002/poi3.198},
	language = {en},
	number = {1},
	urldate = {2026-03-19},
	journal = {Policy \& Internet},
	author = {Wagner, Ben},
	year = {2019},
	note = {\_eprint: https://onlinelibrary.wiley.com/doi/pdf/10.1002/poi3.198},
	pages = {104--122},
}

@inproceedings{alkhatib2019-StreetLevelAlgorithmsTheory,
	address = {New York, NY, USA},
	series = {{CHI} '19},
	title = {Street-{Level} {Algorithms}: {A} {Theory} at the {Gaps} {Between} {Policy} and {Decisions}},
	isbn = {978-1-4503-5970-2},
	shorttitle = {Street-{Level} {Algorithms}},
	url = {https://dl.acm.org/doi/10.1145/3290605.3300760},
	doi = {10.1145/3290605.3300760},
	urldate = {2026-03-18},
	booktitle = {Proceedings of the 2019 {CHI} {Conference} on {Human} {Factors} in {Computing} {Systems}},
	publisher = {Association for Computing Machinery},
	author = {Alkhatib, Ali and Bernstein, Michael},
	month = may,
	year = {2019},
	pages = {1--13},
}

@inproceedings{poe2024-ConflictAlgorithmicFairness,
	address = {New York, NY, USA},
	series = {{FAccT} '24},
	title = {The {Conflict} {Between} {Algorithmic} {Fairness} and {Non}-{Discrimination}: {An} {Analysis} of {Fair} {Automated} {Hiring}},
	isbn = {979-8-4007-0450-5},
	shorttitle = {The {Conflict} {Between} {Algorithmic} {Fairness} and {Non}-{Discrimination}},
	url = {https://dl.acm.org/doi/10.1145/3630106.3659015},
	doi = {10.1145/3630106.3659015},
	urldate = {2026-03-11},
	booktitle = {Proceedings of the 2024 {ACM} {Conference} on {Fairness}, {Accountability}, and {Transparency}},
	publisher = {Association for Computing Machinery},
	author = {Poe, Robert Lee and El Mestari, Soumia Zohra},
	month = jun,
	year = {2024},
	pages = {1907--1916},
}

@inproceedings{balagopalan2022-RoadExplainabilityPaved,
	address = {New York, NY, USA},
	series = {{FAccT} '22},
	title = {The {Road} to {Explainability} is {Paved} with {Bias}: {Measuring} the {Fairness} of {Explanations}},
	isbn = {978-1-4503-9352-2},
	shorttitle = {The {Road} to {Explainability} is {Paved} with {Bias}},
	url = {https://dl.acm.org/doi/10.1145/3531146.3533179},
	doi = {10.1145/3531146.3533179},
	urldate = {2026-03-11},
	booktitle = {Proceedings of the 2022 {ACM} {Conference} on {Fairness}, {Accountability}, and {Transparency}},
	publisher = {Association for Computing Machinery},
	author = {Balagopalan, Aparna and Zhang, Haoran and Hamidieh, Kimia and Hartvigsen, Thomas and Rudzicz, Frank and Ghassemi, Marzyeh},
	month = jun,
	year = {2022},
	pages = {1194--1206},
}

@book{egbert2021-CriminalFuturesPredictive,
	address = {London ; New York},
	title = {Criminal {Futures}: {Predictive} {Policing} and {Everyday} {Police} {Work}},
	isbn = {978-0-367-34926-4},
	shorttitle = {Criminal {Futures}},
	language = {English},
	publisher = {Routledge},
	author = {Egbert, Simon and Leese, Matthias},
	year = {2021},
}

@inproceedings{jacobs2021-MeasurementFairness,
	address = {New York, NY, USA},
	series = {{FAccT} '21},
	title = {Measurement and {Fairness}},
	isbn = {978-1-4503-8309-7},
	url = {https://dl.acm.org/doi/10.1145/3442188.3445901},
	doi = {10.1145/3442188.3445901},
	urldate = {2026-03-10},
	booktitle = {Proceedings of the 2021 {ACM} {Conference} on {Fairness}, {Accountability}, and {Transparency}},
	publisher = {Association for Computing Machinery},
	author = {Jacobs, Abigail Z. and Wallach, Hanna},
	month = mar,
	year = {2021},
	pages = {375--385},
}

@article{amoore2024-DeepBorder,
	title = {The deep border},
	volume = {109},
	issn = {0962-6298},
	url = {https://www.sciencedirect.com/science/article/pii/S0962629821002079},
	doi = {10.1016/j.polgeo.2021.102547},
	urldate = {2026-03-05},
	journal = {Political Geography},
	author = {Amoore, Louise},
	month = mar,
	year = {2024},
	pages = {102547},
}

@article{amoore2006-BiometricBordersGoverning,
	title = {Biometric borders: {Governing} mobilities in the war on terror},
	volume = {25},
	issn = {0962-6298},
	shorttitle = {Biometric borders},
	url = {https://www.sciencedirect.com/science/article/pii/S0962629806000217},
	doi = {10.1016/j.polgeo.2006.02.001},
	number = {3},
	urldate = {2026-03-04},
	journal = {Political Geography},
	author = {Amoore, Louise},
	month = mar,
	year = {2006},
	pages = {336--351},
}

@article{stevens2021-SeeingInfrastructureRace,
	title = {Seeing infrastructure: race, facial recognition and the politics of data},
	volume = {35},
	issn = {0950-2386},
	shorttitle = {Seeing infrastructure},
	url = {https://doi.org/10.1080/09502386.2021.1895252},
	doi = {10.1080/09502386.2021.1895252},
	number = {4-5},
	urldate = {2026-03-04},
	journal = {Cultural Studies},
	publisher = {Routledge},
	author = {Stevens, Nikki and Keyes, Os},
	month = sep,
	year = {2021},
	note = {\_eprint: https://doi.org/10.1080/09502386.2021.1895252},
	pages = {833--853},
}

@article{koepke2018-DangerAheadRisk,
	title = {Danger {Ahead}: {Risk} {Assessment} and the {Future} of {Bail} {Reform}},
	volume = {93},
	shorttitle = {Danger {Ahead}},
	url = {https://digitalcommons.law.uw.edu/wlr/vol93/iss4/4},
	number = {4},
	journal = {Washington Law Review},
	author = {Koepke, John and Robinson, David},
	month = dec,
	year = {2018},
	pages = {1725},
}

@article{green2022-FlawsPoliciesRequiring,
	title = {The flaws of policies requiring human oversight of government algorithms},
	volume = {45},
	issn = {2212473X},
	url = {https://linkinghub.elsevier.com/retrieve/pii/S0267364922000292},
	doi = {10.1016/j.clsr.2022.105681},
	language = {en},
	urldate = {2026-02-21},
	journal = {Computer Law \& Security Review},
	author = {Green, Ben},
	month = jul,
	year = {2022},
	pages = {105681},
}

@article{bovens2002-StreetLevelSystemLevelBureaucracies,
	title = {From {Street}-{Level} to {System}-{Level} {Bureaucracies}: {How} {Information} and {Communication} {Technology} is {Transforming} {Administrative} {Discretion} and {Constitutional} {Control}},
	volume = {62},
	issn = {1540-6210},
	shorttitle = {From {Street}-{Level} to {System}-{Level} {Bureaucracies}},
	url = {https://onlinelibrary.wiley.com/doi/abs/10.1111/0033-3352.00168},
	doi = {10.1111/0033-3352.00168},
	language = {en},
	number = {2},
	urldate = {2026-02-21},
	journal = {Public Administration Review},
	author = {Bovens, Mark and Zouridis, Stavros},
	year = {2002},
	note = {\_eprint: https://onlinelibrary.wiley.com/doi/pdf/10.1111/0033-3352.00168},
	pages = {174--184},
}

@book{flanagan2014-ValuesPlayDigital,
	address = {Cambridge, MA, USA},
	title = {Values at {Play} in {Digital} {Games}},
	isbn = {978-0-262-32445-8},
	language = {en},
	publisher = {MIT Press},
	author = {Flanagan, Mary and Nissenbaum, Helen},
	month = jul,
	year = {2014},
}

@incollection{flanagan2008-EmbodyingValuesTechnology,
	address = {Cambridge},
	series = {Cambridge {Studies} in {Philosophy} and {Public} {Policy}},
	title = {Embodying {Values} in {Technology}: {Theory} and {Practice}},
	isbn = {978-0-521-85549-5},
	shorttitle = {Embodying {Values} in {Technology}},
	url = {https://www.cambridge.org/core/books/information-technology-and-moral-philosophy/embodying-values-in-technology-theory-and-practice/14500603971A7DCF628026A240189CE0},
	doi = {10.1017/CBO9780511498725.017},
	urldate = {2026-01-12},
	booktitle = {Information {Technology} and {Moral} {Philosophy}},
	publisher = {Cambridge University Press},
	author = {Flanagan, Mary and Howe, Daniel C. and Nissenbaum, Helen},
	editor = {van den Hoven, Jeroen and Weckert, John},
	year = {2008},
	pages = {322--353},
}

@article{mulligan-PerfectEnforcementLaw,
	title = {Perfect {Enforcement} {Of} {Law}: {When} {To} {Limit} {And} {When} {To} {Use} {Technology}},
	language = {en},
	number = {4},
	author = {Mulligan, Christina M},
	year = {2008},
}

@misc{-SABRAnalyticsConference,
	title = {{SABR} {Analytics} {Conference}},
	url = {https://sabr.org/analytics},
	language = {en-US},
	urldate = {2026-01-06},
	author = {Society for American Baseball Research},
	year = {2025},
}

@article{dean-UmpiresTechYoure,
	title = {Umpires to {Tech}: {You}'re {Out}!},
	issn = {1059-1028},
	shorttitle = {Umpires to {Tech}},
	url = {https://www.wired.com/2003/06/umpires-to-tech-youre-out/},
	language = {en-US},
	urldate = {2025-12-29},
	journal = {Wired},
	author = {Dean, Katie},
	month = jun,
	year = {2003},
	note = {Section: tags},
}

@inproceedings{srinivasan2024-GeneralizedPeopleDiversity,
	address = {New York, NY, USA},
	series = {{FAccT} '24},
	title = {Generalized {People} {Diversity}: {Learning} a {Human} {Perception}-{Aligned} {Diversity} {Representation} for {People} {Images}},
	isbn = {979-8-4007-0450-5},
	shorttitle = {Generalized {People} {Diversity}},
	url = {https://dl.acm.org/doi/10.1145/3630106.3658940},
	doi = {10.1145/3630106.3658940},
	urldate = {2026-01-07},
	booktitle = {Proceedings of the 2024 {ACM} {Conference} on {Fairness}, {Accountability}, and {Transparency}},
	publisher = {Association for Computing Machinery},
	author = {Srinivasan, Hansa and Schumann, Candice and Sinha, Aradhana and Madras, David and Olanubi, Gbolahan Oluwafemi and Beutel, Alex and Ricco, Susanna and Chen, Jilin},
	month = jun,
	year = {2024},
	pages = {797--821},
}

@book{collins2017-BadCallTechnologys,
	address = {Cambridge, MA, USA},
	series = {Inside {Technology}},
	title = {Bad {Call}: {Technology}’s {Attack} on {Referees} and {Umpires} and {How} to {Fix} {It}},
	isbn = {978-0-262-53444-4},
	shorttitle = {Bad {Call}},
	language = {en},
	publisher = {MIT Press},
	author = {Collins, Harry and Evans, Robert and Higgins, Christopher},
	editor = {Bijker, Wiebe and Jones-Imhotep, Edward and Slayton, Rebecca},
	month = sep,
	year = {2017},
}

@article{mckee2007-JudgesUmpires,
	title = {Judges as {Umpires}},
	volume = {35},
	issn = {00914029},
	url = {https://scholarlycommons.law.hofstra.edu/hlr/vol35/iss4/3},
	number = {4},
	journal = {Hofstra Law Review},
	author = {McKee, Theodore},
	month = jan,
	year = {2007},
}

@article{fried2012-BallsStrikes,
	title = {Balls and {Strikes}},
	volume = {61},
	issn = {0094-4076},
	url = {https://scholarlycommons.law.emory.edu/elj/vol61/iss4/1},
	number = {4},
	journal = {Emory Law Journal},
	author = {Fried, Charles},
	month = jan,
	year = {2012},
	pages = {641},
}

@misc{kerr2010-DigitalLocksAutomation,
	address = {Rochester, NY},
	type = {{SSRN} {Scholarly} {Paper}},
	title = {Digital {Locks} and the {Automation} of {Virtue}},
	url = {https://papers.ssrn.com/abstract=2115655},
	language = {en},
	urldate = {2026-01-07},
	publisher = {Social Science Research Network},
	author = {Kerr, Ian R.},
	month = oct,
	year = {2010},
}

@book{zittrain2008-FutureInternetAndHow,
	address = {New Haven, Conn.},
	title = {The {Future} of the {Internet}--{And} {How} to {Stop} {It}},
	isbn = {978-0-300-15124-4},
	language = {English},
	publisher = {Yale University Press},
	author = {Zittrain, Jonathan},
	year = {2008},
}

@misc{selbst2018-FairnessAbstractionSociotechnical,
	address = {Rochester, NY},
	type = {{SSRN} {Scholarly} {Paper}},
	title = {Fairness and {Abstraction} in {Sociotechnical} {Systems}},
	url = {https://papers.ssrn.com/abstract=3265913},
	language = {en},
	urldate = {2026-01-06},
	publisher = {Social Science Research Network},
	author = {Selbst, Andrew D. and Boyd, Danah and Friedler, Sorelle A. and Venkatasubramanian, Suresh and Janet Vertesi},
	month = aug,
	year = {2018},
}

@article{braun2006-UsingThematicAnalysis,
	title = {Using thematic analysis in psychology},
	volume = {3},
	issn = {1478-0887},
	url = {https://doi.org/10.1191/1478088706qp063oa},
	doi = {10.1191/1478088706qp063oa},
	number = {2},
	urldate = {2026-01-06},
	journal = {Qualitative Research in Psychology},
	publisher = {Routledge},
	author = {Braun, Virginia and Clarke, Victoria},
	month = jan,
	year = {2006},
	note = {\_eprint: https://doi.org/10.1191/1478088706qp063oa},
	pages = {77--101},
}

@article{pinch1984-SocialConstructionFacts,
	title = {The {Social} {Construction} of {Facts} and {Artefacts}: or {How} the {Sociology} of {Science} and the {Sociology} of {Technology} might {Benefit} {Each} {Other}},
	volume = {14},
	issn = {0306-3127},
	shorttitle = {The {Social} {Construction} of {Facts} and {Artefacts}},
	url = {https://doi.org/10.1177/030631284014003004},
	doi = {10.1177/030631284014003004},
	language = {EN},
	number = {3},
	urldate = {2026-01-05},
	journal = {Social Studies of Science},
	publisher = {SAGE Publications Ltd},
	author = {Pinch, Trevor J. and Bijker, Wiebe E.},
	month = aug,
	year = {1984},
	pages = {399--441},
}

@misc{Silver2003-ESPNcomMLBQuesTec,
	title = {{ESPN}.com: {MLB} - {QuesTec} not yet showing consistency from umpires},
	url = {https://www.espn.com/mlb/columns/bp/1563505.html},
	urldate = {2025-12-29},
	author = {Silver, Nate and Woolner, Keith},
	month = jun,
	year = {2003},
}

@article{hunter2018-NewMetricsEvaluating,
	title = {New metrics for evaluating home plate umpire consistency and accuracy},
	volume = {14},
	copyright = {De Gruyter expressly reserves the right to use all content for commercial text and data mining within the meaning of Section 44b of the German Copyright Act.},
	issn = {1559-0410},
	url = {https://www.degruyterbrill.com/document/doi/10.1515/jqas-2018-0061/html},
	doi = {10.1515/jqas-2018-0061},
	language = {en},
	number = {4},
	urldate = {2025-12-29},
	journal = {Journal of Quantitative Analysis in Sports},
	publisher = {De Gruyter},
	author = {Hunter, David J.},
	month = dec,
	year = {2018},
	pages = {159--172},
}

@misc{MLB-PitchtrackingEraGlossary,
	title = {Pitch-tracking {Era} {\textbar} {Glossary}},
	url = {https://www.mlb.com/glossary/miscellaneous/pitch-tracking-era},
	language = {en},
	urldate = {2025-12-29},
	journal = {MLB.com},
	author = {MLB},
}

@book{creswell2011-DesigningConductingMixed,
	address = {Los Angeles},
	title = {Designing and {Conducting} {Mixed} {Methods} {Research}},
	isbn = {978-1-4129-7517-9},
	language = {English},
	publisher = {SAGE Publications, Inc},
	author = {Creswell, John W. and Clark, Vicki L. Plano},
	year = {2011},
}

@article{stark2025-WhatWeLearned,
	title = {What we learned from {MLB}’s spring robot-umpire test: {Players}, managers, execs weigh in},
	issn = {0362-4331},
	shorttitle = {What we learned from {MLB}’s spring robot-umpire test},
	url = {https://www.nytimes.com/athletic/6224826/2025/03/24/mlb-automated-ball-strike-system-spring-training/},
	language = {en-US},
	urldate = {2025-12-26},
	journal = {The New York Times},
	author = {Stark, Jayson},
	month = apr,
	year = {2025},
}

@misc{Maaddi2019-RobotUmpiresDebut,
	title = {'{Robot} umpires' debut in independent {Atlantic} {League}},
	url = {https://apnews.com/article/93f7e8be7f404ff2a6eadf2e9295c9d3},
	language = {en},
	urldate = {2025-12-26},
	journal = {AP News},
	author = {Maaddi, Rob},
	month = jul,
	year = {2019},
}

@article{schmidt2025-QARobManfred,
	chapter = {Briefing},
	title = {A {Q}\&{A} {With} {Rob} {Manfred}, {M}.{L}.{B}.’s {Commissioner}, on the {Future} of {Baseball}},
	issn = {0362-4331},
	url = {https://www.nytimes.com/2025/04/06/briefing/rob-manfred-interview-baseball.html},
	language = {en-US},
	urldate = {2025-12-26},
	journal = {The New York Times},
	author = {Schmidt, Michael S.},
	month = apr,
	year = {2025},
}

@misc{golen2023-RoboUmpsReach,
	title = {Robo umps reach {Triple}-{A}, but {MLB} rollout still uncertain},
	url = {https://apnews.com/article/robo-umps-abs-triplea-ccc901dc69c6101fb6a793e5fe867a77},
	language = {en},
	urldate = {2025-12-26},
	journal = {AP News},
	author = {Golen, Jimmy},
	month = may,
	year = {2023},
}

@misc{Dykstra2020-ToolshedFarmsPrep,
	title = {Toolshed: {Farms} prep for {FSL} robot umps},
	shorttitle = {Toolshed},
	url = {https://www.milb.com/news/toolshed-farm-systems-prepare-for-florida-state-league-robot-umpires-313089276},
	language = {en},
	urldate = {2025-12-26},
	journal = {MiLB.com},
	author = {Dykstra, Sam},
	month = mar,
	year = {2020},
}

@article{stark2024-MLBJustTweaked,
	title = {{MLB} just tweaked {Triple} {A}’s electronic strike zone: {What} you need to know and why it matters},
	issn = {0362-4331},
	shorttitle = {{MLB} just tweaked {Triple} {A}’s electronic strike zone},
	url = {https://www.nytimes.com/athletic/4840104/2023/09/07/triple-a-electronic-strike-zone-changes-mlb-explainer/},
	language = {en-US},
	urldate = {2025-12-26},
	journal = {The New York Times},
	author = {Stark, Jayson},
	month = jul,
	year = {2024},
}

@book{bowker2000-SortingThingsOut,
	address = {Cambridge, MA, USA},
	series = {Inside {Technology}},
	title = {Sorting {Things} {Out}: {Classification} and {Its} {Consequences}},
	isbn = {978-0-262-52295-3},
	shorttitle = {Sorting {Things} {Out}},
	language = {en},
	publisher = {MIT Press},
	author = {Bowker, Geoffrey C. and Star, Susan Leigh},
	editor = {Bijker, Wiebe and Jones-Imhotep, Edward and Slayton, Rebecca},
	month = aug,
	year = {2000},
}

@misc{jedlovec2020-IntroducingStatcast2020,
	title = {Introducing {Statcast} 2020: {Hawk}-{Eye} and {Google} {Cloud}},
	shorttitle = {Introducing {Statcast} 2020},
	url = {https://technology.mlblogs.com/introducing-statcast-2020-hawk-eye-and-google-cloud-a5f5c20321b8},
	language = {en},
	urldate = {2025-12-17},
	journal = {Medium},
	author = {Jedlovec, Ben},
	month = jul,
	year = {2020},
}

@book{zumsteg2007-CheatersGuideBaseball,
	address = {Boston},
	title = {The {Cheater}'s {Guide} {To} {Baseball}: {A} {Lively} {History} of {Sign} {Stealing}, {Scandals}, and the {Game}'s {Greatest} {Manipulators}},
	isbn = {978-0-618-55113-2},
	shorttitle = {The {Cheater}'s {Guide} {To} {Baseball}},
	language = {English},
	publisher = {Mariner Books},
	author = {Zumsteg, Derek},
	year = {2007},
}

@book{hershberger2019-StrikeFourEvolution,
	address = {Lanham, Maryland},
	title = {Strike {Four}: {The} {Evolution} of {Baseball}},
	isbn = {978-1-5381-2114-6},
	shorttitle = {Strike {Four}},
	language = {English},
	publisher = {Rowman \& Littlefield Publishers},
	author = {Hershberger, Richard and Thorn, John},
	year = {2019},
}

@book{lam2025-FewerRulesBetter,
	title = {Fewer {Rules}, {Better} {People}: {The} {Case} for {Discretion}},
	isbn = {978-1-324-05124-4},
	language = {en},
	publisher = {W. W. Norton \& Company},
	author = {Lam, Barry},
	month = feb,
	year = {2025},
}

@book{levy2022-DataDrivenTruckers,
	title = {Data {Driven}: {Truckers}, {Technology}, and the {New} {Workplace} {Surveillance}},
	isbn = {978-0-691-17530-0},
	shorttitle = {Data {Driven}},
	language = {en},
	publisher = {Princeton University Press},
	author = {Levy, Karen},
	month = dec,
	year = {2022},
	note = {Publisher: Princeton University Press},
}

@book{schauer1991-PlayingRulesPhilosophical,
	title = {Playing by the {Rules}: {A} {Philosophical} {Examination} of {Rule}-{Based} {Decision}-{Making} in {Law} and in {Life}},
	isbn = {978-0-19-101874-9},
	shorttitle = {Playing by the {Rules}},
	language = {en},
	publisher = {Clarendon Press},
	author = {Schauer, Frederick},
	month = aug,
	year = {1991},
	note = {Google-Books-ID: CNXhBQAAQBAJ},
}

@misc{cooper2025-DitchingMLBRule,
	title = {By {Ditching} {MLB} {Rule} {Book}, {ABS} {Strike} {Zone} {Has} {Found} {Its} {Footing}},
	url = {https://www.baseballamerica.com/stories/by-ditching-mlb-rule-book-abs-strike-zone-has-found-its-footing/},
	language = {en-US},
	urldate = {2025-12-12},
	journal = {College Baseball, MLB Draft, Prospects - Baseball America},
	author = {Cooper, J. J.},
	month = mar,
	year = {2025},
}

@inproceedings{kamino2025-MillionEyesRobot,
	address = {Melbourne, Australia},
	series = {{HRI} '25},
	title = {Million {Eyes} on the '{Robot} {Umps}': {The} {Case} for {Studying} {Sports} in {HRI} {Through} {Baseball}},
	shorttitle = {Million {Eyes} on the '{Robot} {Umps}'},
	urldate = {2025-12-09},
	booktitle = {Proceedings of the 2025 {ACM}/{IEEE} {International} {Conference} on {Human}-{Robot} {Interaction}},
	publisher = {IEEE Press},
	author = {Kamino, Waki and Wen-Yi, Andrea W. and Agarwal, Dhruv and Hamilton, Sil and Kang, Eun Jeong and Kim, Jieun and Kusumegi, Keigo and Moradi, Pegah and Mwesigwa, Daniel and Tao, Yan and Tsai, I-Ting and Yang, Ethan and Zhu, Shengqi and Han, Shu-Jung and Lee, Chi-Jung and Sack, Michael Joseph and Yu, Tianhong Catherine and Khoo, Weslie and Ricci, Andy Elliot and Hou, Yoyo Tsung-Yu and Kim, Boyoung and šabanović, Selma and Crandall, David J. and Levy, Karen and Jung, Malte F.},
	month = mar,
	year = {2025},
	pages = {1383--1388},
}

@misc{shein2024-AIJudgingSports,
	title = {{AI} {Judging} in {Sports} – {Communications} of the {ACM}},
	url = {https://cacm.acm.org/news/ai-judging-in-sports/},
	language = {en-US},
	urldate = {2024-11-27},
	author = {Shein, Esther},
	month = nov,
	year = {2024},
}

@misc{jones2018-SportingChancesRobot,
	address = {Rochester, NY},
	type = {{SSRN} {Scholarly} {Paper}},
	title = {Sporting {Chances}: {Robot} {Referees} and the {Automation} of {Enforcement}},
	shorttitle = {Sporting {Chances}},
	url = {https://papers.ssrn.com/abstract=3293076},
	doi = {10.2139/ssrn.3293076},
	language = {en},
	urldate = {2025-12-09},
	publisher = {Social Science Research Network},
	author = {Jones, Meg Leta and Levy, Karen},
	month = nov,
	year = {2018},
}

@article{kiviat2025-ExceptionsAlgorithmicAge,
	title = {Exceptions in the {Algorithmic} {Age}: {Evidence} from the {Case} of {Tenant} {Screening}},
	issn = {0002-9602},
	shorttitle = {Exceptions in the {Algorithmic} {Age}},
	url = {https://www.journals.uchicago.edu/doi/10.1086/739108},
	doi = {10.1086/739108},
	urldate = {2025-12-09},
	journal = {American Journal of Sociology},
	publisher = {The University of Chicago Press},
	author = {Kiviat, Barbara and Greene, Sara  Sternberg and Yoon, Hesu},
	month = oct,
	year = {2025},
}

@article{rich2012-ShouldWeMake,
	title = {Should {We} {Make} {Crime} {Impossible}?},
	issn = {1556-5068},
	url = {http://www.ssrn.com/abstract=2029201},
	doi = {10.2139/ssrn.2029201},
	language = {en},
	urldate = {2025-10-24},
	journal = {SSRN Electronic Journal},
	author = {Rich, Michael},
	year = {2012},
}

@online{atlanta,
          title = {Southern Umpires Camp},
          author = {Southern and International Baseball Umpire Training Camps},
          year = {2025},
          url = {https://baseballumpirecamps.com/page/southern},
          urldate = {2025}
        }

@misc{wikimedia_morichan,
  author = {Mori Chan},
  title = {Explanation of the strike zone in baseball.},
  year = {2008},
  howpublished = {Wikimedia Commons},
  note = {Licensed under CC BY 2.0},
  url = {https://commons.wikimedia.org/w/index.php?curid=5591282}
}

@online{dallas,
          title = {MLB Umpire Camps},
          author = {{MLB Advanced Media, LP.}},
          year = {2025},
          url = {https://www.mlb.com/official-information/umpires/camps},
          urldate = {2025}
        }

@book{hammersley2019ethnography,
  title={Ethnography: Principles in practice},
  author={Hammersley, Martyn and Atkinson, Paul},
  year={2019},
  publisher={Routledge}
}

@article{pound1910law,
  title={Law in books and law in action},
  author={Pound, Roscoe},
  journal={American Law Review},
  volume={44},
  pages={12},
  year={1910},
  publisher={HeinOnline}
}

@book{lessig2009code,
  title={Code: And other laws of cyberspace},
  author={Lessig, Lawrence},
  year={2009},
  publisher={ReadHowYouWant. com}
}

@article{kleinberg2016inherent,
  title={Inherent trade-offs in the fair determination of risk scores},
  author={Kleinberg, Jon and Mullainathan, Sendhil and Raghavan, Manish},
  journal={arXiv preprint arXiv:1609.05807},
  year={2016}
}

\appendix

\end{document}